\documentclass[12pt]{iopart}

\usepackage{graphicx} 
\usepackage[colorlinks=true,linkcolor=blue,citecolor=red]{hyperref}%
\usepackage{cite}
\usepackage{amssymb}
\usepackage{bm}

\usepackage{tikz-cd}

\def\Pr{\mathcal{P}}
\def\L{\mathcal{L}}
\def\N{\mathcal{N}}
\def\P{\mathbb{P}}
\def\T{\mathcal{T}}

\def\Ti{\T^0}
\def\I{\mathbf{1}}

\def\x{\bm{x}}
\def\X{\bm{X}}
\def\R{\mathbb{R}}

\def\koff{k_{\rm off}}
\def\kon{k_{\rm on}}
\def\pa{\partial\Omega}

\def\H{\mathcal{H}}

\def\erfc{\mathrm{erfc}}
\def\erfcx{\mathrm{erfcx}}

\def\ctan{\mathrm{ctan}}

\begin{document}

\title[First-passage times of multiple particles]{First-passage times of multiple diffusing particles with reversible target-binding kinetics}

\author{Denis~S.~Grebenkov}
 \ead{denis.grebenkov@polytechnique.edu}
\address{
$^1$ Laboratoire de Physique de la Mati\`{e}re Condens\'{e}e (UMR 7643), \\ 
CNRS -- Ecole Polytechnique, IP Paris, 91120 Palaiseau, France}

\address{
$^2$ Institute for Physics and Astronomy, University of Potsdam, 14476
Potsdam-Golm, Germany}

\author{Aanjaneya Kumar}
  \ead{kumar.aanjaneya@students.iiserpune.ac.in}
\address{
Department of Physics, Indian Institute of Science Education and Research, \\ Dr. Homi Bhabha Road, Pune 411008, India}

\date{\today}

\begin{abstract}
We investigate a class of diffusion-controlled reactions that are
initiated at the time instance when a prescribed number $K$ among $N$
particles independently diffusing in a solvent are simultaneously
bound to a target region.  In the irreversible target-binding setting,
the particles that bind to the target stay there forever, and the
reaction time is the $K$-th fastest first-passage time to the target,
whose distribution is well-known.  In turn, reversible binding, which
is common for most applications, renders theoretical analysis much
more challenging and drastically changes the distribution of reaction
times.  We develop a renewal-based approach to derive an approximate
solution for the probability density of the reaction time.  This
approximation turns out to be remarkably accurate for a broad range of
parameters.  We also analyze the dependence of the mean reaction time
or, equivalently, the inverse reaction rate, on the main parameters
such as $K$, $N$, and binding/unbinding constants.  Some biophysical
applications and further perspectives are briefly discussed.
\end{abstract}

\pacs{02.50.-r, 05.40.-a, 02.70.Rr, 05.10.Gg}



\noindent{\it Keywords\/}: First-passage time, Diffusion-controlled reactions, Reversible binding, Extreme statistics

\submitto{\JPA}

\maketitle

\section{Introduction}

Diffusion-controlled processes and reactions play the central role in
microbiology, physiology and many industrial procedures
\cite{Lauffenburger,Alberts,Redner,Schuss,Metzler,Oshanin,Grebenkov07,Benichou14,Holcman14}.
In a common setting of bimolecular reactions, two particles (e.g., a
ligand and a receptor) need to meet each other to initiate a reaction
event.  As the encounter results from the stochastic motion of one or
both particles, the reaction time is random.  Since the seminal work
by von Smoluchowski \cite{Smoluchowski1917}, such first-encounter or
first-passage problems have been thoroughly investigated.  Among
various studied aspects, one can mention the impact of stuctural
organization and dynamical heterogeneities of the medium
\cite{Condamin07,Benichou10b,Ghosh16,Grebenkov16,Levernier19,Grebenkov19b,Hartich19,Hartich19b,Grebenkov20b},
the asymptotic behavior of the reaction rate and the mean
first-passage time in the small-target limit
\cite{Grigoriev02,Singer06a,Singer06b,Singer06c,Benichou08,Pillay10,Cheviakov10,Cheviakov12,Caginalp12,Mattos12,Berezhkovsky12,Marshall16,Grebenkov17b},
distinct features of the whole distribution
\cite{Rupprecht15,Godec16,Godec16b,Grebenkov18a,Grebenkov18,Grebenkov19},
and the effect of target mobility
\cite{Redner99,Oshanin09,Lawley19b,LeVot20,LeVot22}.

However, there exist more sophisticated processes (that we will still
call ``reactions'') involving multiple particles.  In microbiology,
there are many activation mechanisms controlled by a threshold
crossing such as signalling in neurons, synaptic plasticity, cell
apoptosis caused by double strand DNA breaks, cell differentiation and
division \cite{Wolpert96,Reddy07,DaoDuc10}.  For instance, binding of
five calcium ions to a calcium-ion-sensing protein initiates a release
of neurotransmitters in the signalling process between two neurons
\cite{Berridge03,Eggermann12,Nakamura15,Dittrich13,Guerrier16,Reva21}.  
Similarly, the ryanodine receptor is activated when two calcium ions
bind to the receptor binding sites \cite{Basnayake19}.  In these
examples, the biochemical event such as signal transmission starts
when a fixed number $K$ among $N$ diffusing particles are
simultaneously bound to the target region for the first time.  If
$\N(t)$ denotes the number of bound particles at time $t$, the
reaction time $\T_{K,N} = \inf\{ t>0 ~:~ \N(t) = K\}$ is the
first-crossing time of a fixed threshold $K$ by the stochastic
non-Markovian process $\N(t)$.  In the idealized case of irreversible
binding when any particle after its binding to the target stays bound
forever, this is the problem of finding the $K$-th fastest
first-passage time $\T^0_{K,N}$ to the target
\cite{Weiss83,Basnayake18,Basnayake19,Schuss19,Lawley20a,Lawley20b,Grebenkov20,Madrid20b,Majumdar20}.
If the particles diffuse independently, the distribution of
$\T^0_{K,N}$ can be easily expressed in terms of the survival
probability for a single particle (see \ref{sec:cases}).  In most
cases, however, binding is reversible so that some particles can
unbind and resume their diffusion before the binding of the $K$-th
fastest particle that renders the problem of such ``impatient''
particles \cite{Grebenkov17} much more challenging.  Recently, Lawley
and Madrid proposed an elegant approximation, in which the
first-binding time and the rebinding time $\tau$ after each unbinding
event were assumed to obey an exponential law.  The process $\N(t)$
could thus be approximated by a Markovian birth-death process, for
which the distribution of the first-crossing time is known explicitly
\cite{Lawley19} (see also \cite{DaoDuc10}).  In the special case $K =
N$, we derived the exact solution of the problem of impatient
particles and showed both advantages and limitations of the
Lawley-Madrid approximation (LMA) \cite{Grebenkov22}.  Despite its
crucial role in providing us with analytical insight into the problem
of impatient particles and the validity of its approximate treatments,
the case when all particles have to bind the target is not so common
in applications.

In this paper, we investigate the general problem of impatient
particles in a common setting when all particles start from
independent uniformly distributed positions.  First, we revisit the
Lawley-Madrid approximation and discuss its validity range.  In
particular, we argue that the key assumption of the LMA requires that
the target is small {\it and} weakly reactive.  The condition of weak
reactivity, which was not emphasized on in \cite{Lawley19}, limits the
applicability of this approximation.  To overcome this limitation, we
develop an alternative approach to the general problem.  Our
approximate solution is confronted to Monte Carlo simulations and
shown to be remarkably accurate for a broad range of parameters.  It
allowed us to investigate the short-time and long-time behaviors of
the probability density of the reaction time $\T_{K,N}$, the
dependence of the mean reaction time on the unbinding rate, and the
role of the numbers $K$ and $N$.

The paper is organized as follows.  In Sec. \ref{sec:problem}, we
formulate the problem of impatient particles and discuss the LMA.
Section \ref{sec:solution} presents the main steps of our approach and
summarizes the approximate formulas for the probability density of the
reaction time $\T_{K,N}$, its short-time and long-time behaviors, and
the mean reaction time.  In Sec. \ref{sec:discussion}, we illustrate
these results for an emblematic model of restricted diffusion between
concentric spheres.  We discuss the accuracy of our approximation and
its limitations.  Section \ref{sec:conclusion} concludes the paper and
suggests further perspectives.  As our derivations are technically
elaborate, most mathematical details are re-delegated to Appendices in
order to facilitate the main text for a wider audience.

\section{Problem of impatient particles}
\label{sec:problem}

\begin{figure}
\begin{center}
\includegraphics[width=65mm]{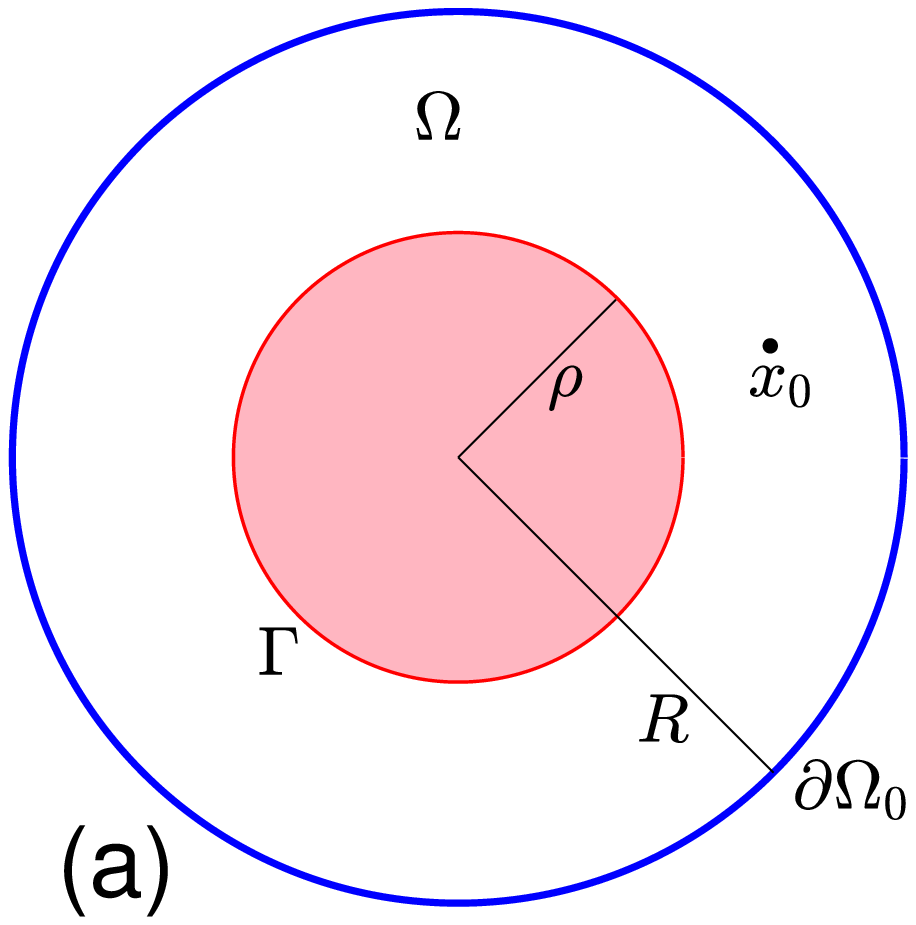} 
\includegraphics[width=85mm]{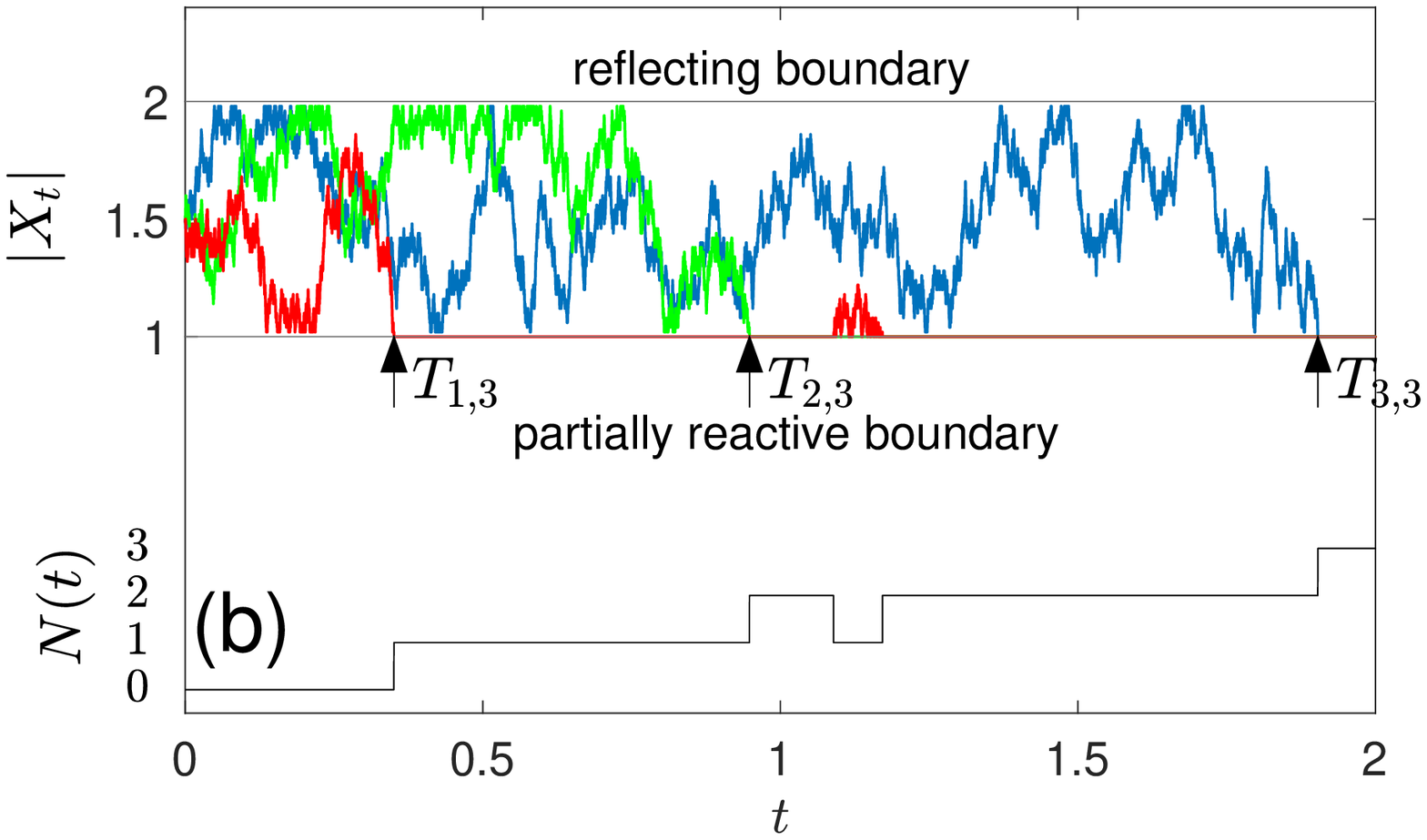} 
\end{center}
\caption{
{\bf (a)} A planar illustration of a bounded domain $\Omega$ between
two concentric spheres of radii $\rho = 1$ and $R = 2$, whose disjoint
boundary $\pa = \pa_0 \cup \Gamma$ is composed of the reflecting outer
sphere $\pa_0$ and the partially reactive inner sphere $\Gamma$.  {\bf
(b)} A numerical simulation for three diffusing particles.  Upper plot
shows the radial coordinate, $|\X_t|$, of simulated trajectories of
three particles that start from a fixed initial position with $|\x_0|
= 1.5$ and diffuse independently, with eventual bindings to the
target.  Arrows indicate the first-crossing times $\T_{1,3}$,
$\T_{2,3}$, and $\T_{3,3}$.  Bottom plot illustrates the number of
bound particles at time $t$, $\N(t)$.  At the beginning, all three
particles are free, and $\N(0) = 0$.  At $\T_{1,3}$, the ``red''
particle binds, switching the counter $\N(t)$ to $1$.  At $\T_{2,3}$,
the ``green'' particle binds, switching the counter $\N(t)$ to $2$.
Few moments later, the ``red'' particle unbinds, diffuses and rebinds
to the target.  Finally, the last ``blue'' particle binds at time
$\T_{3,3}$, switching the counter $\N(t)$ to $3$.}
\label{fig:scheme}
%
\end{figure}

We consider $N$ particles that independently diffuse with diffusion
coefficient $D$ inside a bounded domain $\Omega \subset \R^d$ with a
smooth boundary $\pa$ that is reflecting everywhere except for a
target region $\Gamma$ with a finite reactivity $\kappa$.  For
instance, $\Omega$ may represent the cytoplasm of a living cell,
surrounded by a plasma membrane $\pa$ that is impermeable for
diffusing particles, and $\Gamma$ be the boundary of an organelle or a
sensor protein on that membrane.  The reactivity $\kappa$ (in
units m/s) is related to the binding probability and characterizes how
easily the particle can bind the target upon their encounter, ranging
from $\kappa = 0$ for an inert target (no binding) to $\kappa =
\infty$ for a perfectly reactive target (binding upon the first
encounter).  The finite reactivity may represent the effect of an
energetic or entropic barrier for binding, stochastic switching
between open and closed states of the target (e.g., an ion channel),
microscopic heterogeneity of the target, etc.
\cite{Collins49,Sano79,Shoup82,Zwanzig90,Sapoval94,Filoche99,Benichou00,Grebenkov03,Berezhkovskii04,Grebenkov06,Grebenkov06a,Reingruber09,Grebenkov10,Lawley15,Bernoff18b}.
In (bio)chemistry, the reactivity is usually expressed in terms of the
forward (bimolecular) reaction rate $\kon$ via $\kappa =
\kon/(|\Gamma|N_A)$, where $|\Gamma|$ is the surface area of the
target and $N_A$ is the Avogadro number \cite{Lauffenburger}.  After
binding, each particle stays on the target region for a random
exponentially distributed waiting time, characterized by the unbinding
rate $\koff$, and then resumes its diffusion from a uniformly
distributed point on $\Gamma$.  The particle diffuses in $\Omega$
until the next binding, and so on (Fig. \ref{fig:scheme}).  In other
words, each particle alternates between free and bound states.  We aim
at describing the random reaction time $\T_{K,N}$, i.e., the first
instance when $K$ particles among $N$ are simultaneously in the bound
state on the target region that is considered as a trigger of the
underlying biochemical process (a reaction event).  As binding and
unbinding events of all particles are independent from each other and
thus asynchronized, finding the probability density $\H_{K,N}(t)$ of
the $\T_{K,N}$ is a challenging open problem.  Note that the above
problem of impatient particles resembles some stochastic models of
multi-channel particulate transport with blockage
\cite{Barre13,Barre15,Barre18}.

The first-binding time $\tau_0$ and the consequent rebinding times
$\tau_1, \tau_2, \ldots$ of any particle are random variables, which
are characterized by the survival probabilities $S(t|\x_0) =
\P_{\x_0}\{\tau_0 > t\}$ and $S(t) = \P\{\tau_i > t\}$, where $\x_0$
is the starting point of the particle, and $\P\{\ldots\}$ denotes the
probability of a random event between braces.  Lawley and Madrid
proposed a remarkable approximation, which relied on the approximation
of these probabilities by an exponential function:
\begin{equation}  \label{eq:LMA_approx}
S(t|\x_0) \approx S(t) \approx e^{-\nu t},
\end{equation}
with an appropriate rate $\nu$ \cite{Lawley19}.  They argued that this
approximation is valid for any small and/or weakly reactive target
such that
\begin{equation}  \label{eq:LMA_cond}
\epsilon = \frac{\kappa \,|\Gamma|\, |\Omega|}{D|\pa|^2 } \ll 1,
\end{equation}
where $|\Omega|$ is the volume of the confining domain, $|\pa|$ and
$|\Gamma|$ are the surface areas of the whole boundary and of the
target region (a reactive subset of $\pa$), respectively.  For
clarity, we focus here on a three-dimensional setting, $d = 3$, but
the arguments are valid in higher dimensions as well.  In
\ref{sec:LMA}, we summarize the explicit formulas of the LMA and
discuss the validity of the condition (\ref{eq:LMA_cond}), which
actually combines two distinct properties of the target: its relative
size and reactivity.  We argue that the LMA is applicable when the
target is small {\it and} weakly reactive.  For instance, when the
target is a sphere of radius $\rho$, the following two conditions
should be fulfilled:
\begin{equation}  \label{eq:LMA_cond2}
\rho \ll R = \frac{|\pa|^2}{4\pi |\Omega|} \,, \qquad \frac{\kappa \rho}{D} \ll 1.
\end{equation}
The first condition is purely geometrical (smallness of the target as
compared to the confining domain), while the second condition involves
both the reactivity and the size of the target but does not depend on
the confining domain.  These two conditions evidently imply
Eq. (\ref{eq:LMA_cond}), but the opposite claim is not true.  In
particular, if the target is small but highly reactive, the second
condition may not be valid, even if Eq. (\ref{eq:LMA_cond}) is
fulfilled.  This situation will be illustrated in
Sec. \ref{sec:discussion}.

\section{Approximate solution}
\label{sec:solution}

To overcome the constraint on weak reactivity, we develop an
alternative approach, which does not rely on the approximation
(\ref{eq:LMA_approx}).  For this purpose, we extend the derivation in
Ref. \cite{Grebenkov22} that was specific to the case $K = N$ and
based on a renewal-type equation
\begin{equation}  \label{eq:PN0}
\Pr_t(N|0) = \int\limits_0^t dt' \, \H_{N,N}(t') \, \Pr_{t-t'}(N|N),
\end{equation}
where $\Pr_t(m|n)$ is the probability of transition from a state with
$n$ bound particles to a state with $m$ bound particles.  Expressing
both $\Pr_t(N|0)$ and $\Pr_{t-t'}(N|N)$ in terms of known occupation
probabilities for a single particle and applying the Laplace transform
led to the probability density $\H_{N,N}(t)$ in the Laplace domain.

A direct extension of this equation to the general case $K < N$ fails.
In fact, the probability $\Pr_t(K|0)$ can still be expressed as an
integral of $\H_{K,N}(t')$ with the probability $\Pr_{t-t'}(K|K)$ of
transition from a state with $K$ bound particles to another state with
$K$ bound particles.  However, this probability also depends on {\it
random} positions of the remaining $N-K$ free particles at time $t'$
that should be averaged out.  Even for independently diffusing
particles, an exact computation of this average remains an open
problem (see \ref{sec:distribution} for further discussion).
Moreover, the resulting probability would be a function of both $t-t'$
and $t'$ so that an extension of Eq. (\ref{eq:PN0}) would be no longer
a convolution, and thus would not be simplified in the Laplace domain.

This fundamental difficulty can be partly resolved in the case when
the starting positions of $N$ particles are uniformly distributed in
the confining domain.  The key point is that the distribution of any
{\it free} particle that started uniformly remains to be almost
uniform at all times, except for a boundary layer near the target
region.  When the target is small and not too highly reactive, this
boundary layer is narrow and can be neglected so that all free
particles can be approximately treated as uniformly distributed at any
time $t'$.  As a consequence, the average of $\Pr_{t-t'}(K|K)$ turns
out to be only a function of $t-t'$, thus keeping the convolution form
of the renewal equation:
\begin{equation}  \label{eq:PNK}
\Pr_t(K|0) = \int\limits_0^t dt' \, H_{K,N}(t') \, \overline{\Pr_{t-t'}(K|K)} ,
\end{equation}
where overline denotes the average over the uniform positions of $N-K$
free particles.  In other words, this integral equation determines an
approximation $H_{K,N}(t)$ of the probability density $\H_{K,N}(t)$ of
the reaction time $\T_{K,N}$.  Both transition probabilities in
Eq. (\ref{eq:PNK}) can be found using combinatorial arguments, namely,
\begin{equation}   \label{eq:PrK0}
\Pr_t(K|0) = {N \choose K} [P(t|\circ)]^K [1 - P(t|\circ)]^{N-K}
\end{equation}
and
\begin{eqnarray}  \nonumber
\overline{\Pr_t(K|K)} &=& \sum\limits_{j=0}^K {K\choose j} [Q(t)]^{K-j} [1-Q(t)]^{j} {N-K \choose j} \\    \label{eq:PrKK}
&\times&  [P(t|\circ)]^{j} [1 - P(t|\circ)]^{N-K-j} ,
\end{eqnarray}
where we use the convention for binomial coefficients that ${n\choose
k} = 0$ for $n<k$.  Here $P(t|\circ)$ (resp., $Q(t)$) is the
probability of finding a particle that was free with uniform initial
distribution (resp., bound) at time $0$, in the bound state at time
$t$.  For instance, the term with $j=0$ in Eq. (\ref{eq:PrKK})
describes the configuration when all $K$ initially bound particles are
found to be bound at time $t$ (note that they can unbind and rebind in
the meantime), while $N-K$ initially free particles are found to be
free at time $t$ (they can also bind and unbind in the meantime).
Similarly, the term with $j = 1$ describes the configuration when
$K-1$ initially bound particles are found to be bound at time $t$, one
initially bound particle is found to be free at time $t$, $N-K-1$
initially free particles are found to be free at time $t$, while one
initially free particle is found to be bound at time $t$ (and all
these particles can undertake an arbitrary number of binding/unbinding
events in the meantime).  In \ref{sec:single}, we show that
\begin{equation}  \label{eq:Pcirc}
P(t|\circ) = \frac{1-Q(t)}{\koff \langle\tau\rangle} \,, 
\end{equation}
whereas $Q(t)$ can be expressed in terms of the probability density
$H(t)$ of the rebinding time for a single particle, and
$\langle\tau\rangle$ is the mean rebinding time.  In
\cite{Grebenkov22}, we derived a very simple and general expression
for this quantity:
\begin{equation}  \label{eq:tau}
\langle\tau\rangle = \frac{|\Omega|}{\kappa |\Gamma|} = \frac{N_A |\Omega|}{\kon}
\end{equation}
(we reproduce its derivation in \ref{sec:distribution}).  Here, it is
expressed in terms of the volume $|\Omega|$ of the confining domain,
the surface area $|\Gamma|$ of the target region, and its reactivity
$\kappa$ or, equivalently, in terms of the forward reaction constant
$\kon$.  Counter-intuitively, the mean rebinding time does not depend
on the diffusion coefficient $D$.  This is a particular example of the
invariance property of general random walks in bounded domains that
the mean traveled distance (and thus the mean exit time) does not
depend on the dynamics of the diffusing particles that enter and exit
the domain through the same subset of the boundary (here, the target)
\cite{Blanco03,Mazzolo04,Benichou05,Mazzolo09}.
Solving the convolution equation (\ref{eq:PNK}) in the Laplace domain,
we obtain the approximate probability density $H_{K,N}(t)$ of the
reaction time $\T_{K,N}$:
\begin{equation}  \label{eq:HKN_exact}
H_{K,N}(t) = \L^{-1} \left\{ \frac{\L\{\Pr_t(K|0)\}}{\L\{\overline{\Pr_t(K|K)}\}} \right\} ,
\end{equation}
where $\L$ and $\L^{-1}$ denote respectively the forward and inverse
Laplace transforms.  This approximate solution of the general problem
of impatient particles constitutes the main result of the paper.  For
$K = N$, one has $\Pr_t(K|0) = [P(t|\circ)]^N$ and
$\overline{\Pr_t(K|K)} = [Q(t)]^N$ and thus retrieves an extension of
the exact solution from Ref. \cite{Grebenkov22} to the case of the
uniform initial distribution of the particles.

In addition to a direct numerical way of computing the approximate
probability density $H_{K,N}(t)$ (see \ref{sec:computation} for
details), Eq. (\ref{eq:HKN_exact}) opens a way to access the
short-time and long-time asymptotic behaviors of this density (see
\ref{sec:asymptotics}):
\begin{eqnarray}  \label{eq:HKN_short}
H_{K,N}(t) &\approx&  K {N\choose K} \, \frac{t^{K-1}}{\langle\tau\rangle^K}  \qquad (t\to 0), \\  \label{eq:HKN_long}
H_{K,N}(t) &\propto& \exp(-t/T_{K,N}) \qquad (t\to\infty),
\end{eqnarray}
where $T_{K,N}$ is the decay time whose approximation reads
\begin{equation}  \label{eq:TKN_decay}
T_{K,N} \approx \frac{1}{\overline{\Pr_{\infty}(K|K)}} \int\limits_0^\infty dt \biggl(\overline{\Pr_t(K|K)} - \overline{\Pr_{\infty}(K|K)}\biggr),
\end{equation}
in which $\overline{\Pr_{\infty}(K|K)}$ is given by
Eq. (\ref{eq:PrKK}) with $P(\infty|\circ) = Q(\infty) = 1/(1 + \koff
\langle\tau\rangle)$.  In addition, our approximate solution allows us to
evaluate the moments of the reaction time $\T_{K,N}$.  For instance,
we derived the following approximation for the mean reaction time (see
\ref{sec:mean})
\begin{equation}   \label{eq:Tmean}
\langle \T_{K,N}\rangle \approx \frac{1}{\Pr_\infty(K|0)} \int\limits_0^\infty dt \biggl(\overline{\Pr_{t}(K|K)} - 
\Pr_{t}(K|0)\biggr) .
\end{equation}
Note that this expression is similar to Eq. (\ref{eq:TKN_decay}) for
the decay time, and they usually yield very close results.
 
The dimensionless parameter $\eta = \koff \langle\tau\rangle \propto
\koff/\kon$ determines whether the reversible binding kinetics is
relevant ($\eta \gtrsim 1$) or not ($\eta \ll 1$).  As discussed in
\ref{sec:mean}, Eq. (\ref{eq:Tmean}) fails as $\eta \to 0$
but gets more and more accurate as $\eta$ increases.  For $\eta \gg
1$, the integral in Eq. (\ref{eq:Tmean}) can be approximately
evaluated as
\begin{equation}  \label{eq:TKN_eta}
\langle \T_{K,N}\rangle  \approx \langle\tau\rangle \, \frac{(\koff\langle\tau\rangle)^{K-1}}{K {N \choose K}}  \qquad (\eta \gg 1).
\end{equation}
For $K = 1$, the approximate mean reaction time $\langle
\T_{1,N}\rangle \approx \langle\tau\rangle/N$ does not depend on
$\koff$, as the first-binding event is independent of the unbinding
kinetics.  This mean value decreases inversely proportional to $N$, as
discussed earlier in Ref. \cite{Grebenkov20,Madrid20b} in the context
of the fastest first-passage time problem.  In the case $K\ll N$, the
above expression reads
\begin{equation}
\langle \T_{K,N}\rangle  \approx \koff (K-1)! \left(\frac{\kappa |\Gamma| N}{\koff |\Omega|}\right)^K ,
\end{equation}
which resembles the asymptotic behavior of the mean first-passage time
of a rare event that $K$ among $N$ independent random walkers
accumulate at a given site of a lattice \cite{Sanders08}.

\section{Discussion}
\label{sec:discussion}

\begin{figure}
\begin{center}
\includegraphics[width=77mm]{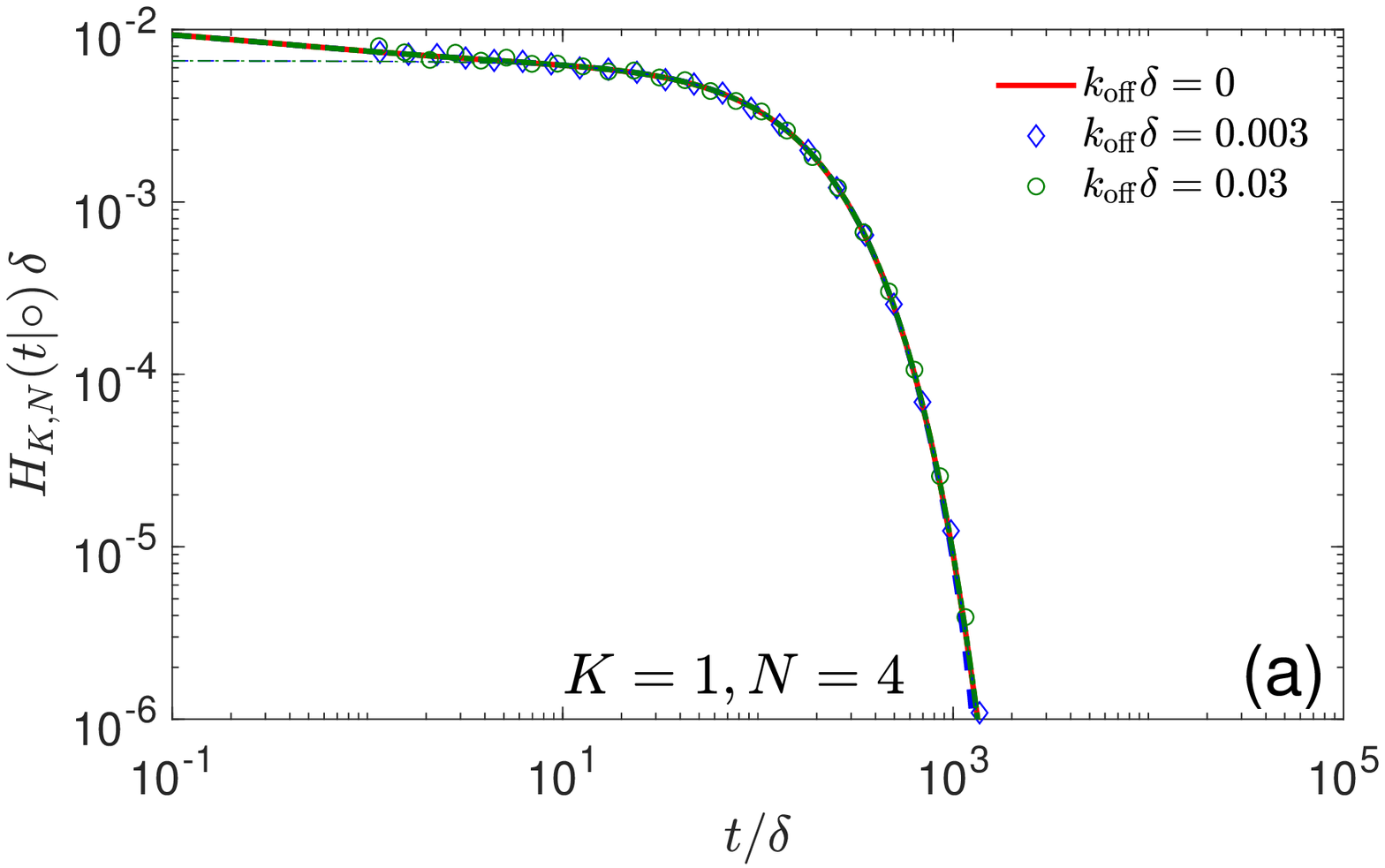}  
\includegraphics[width=77mm]{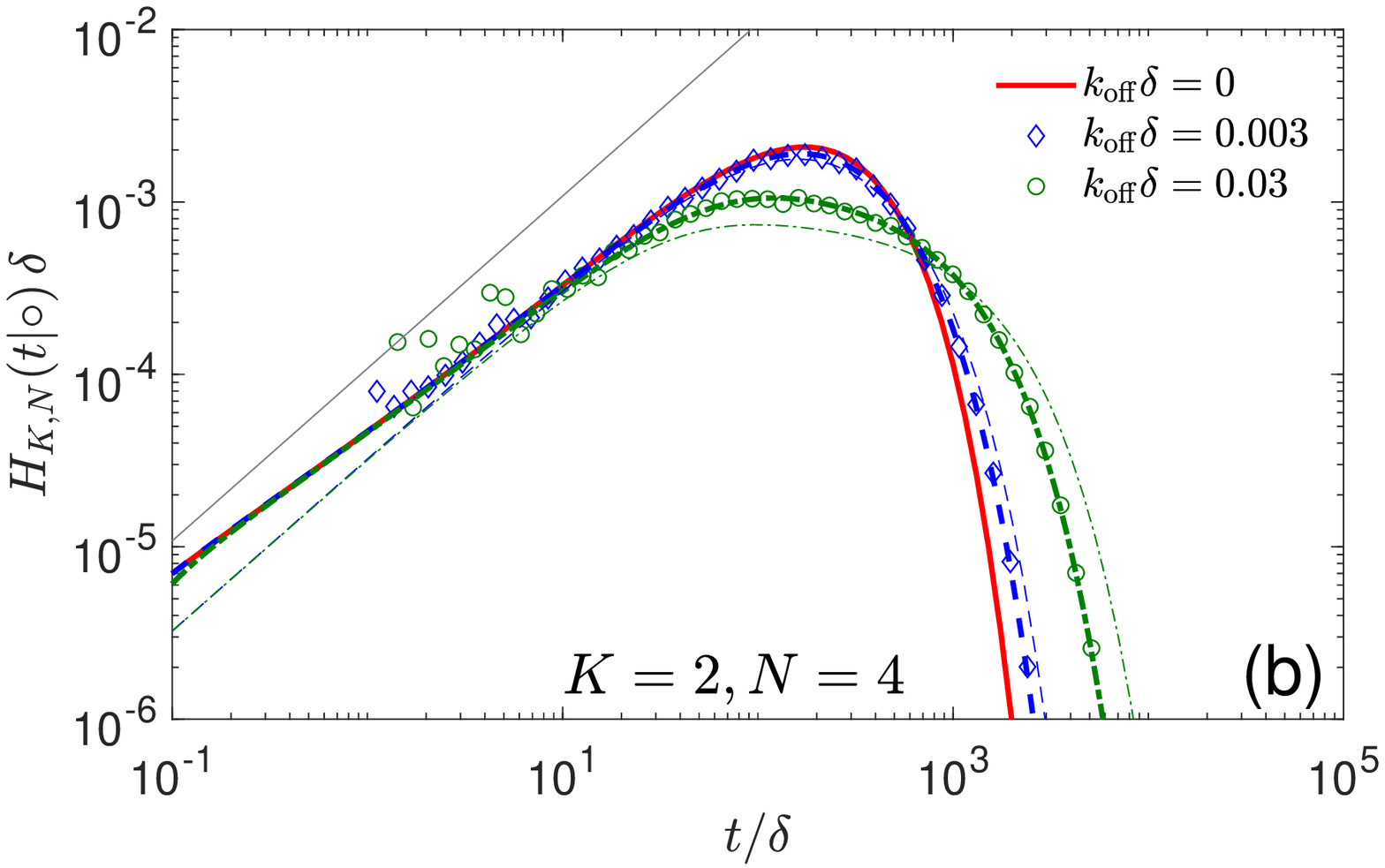}  
\includegraphics[width=77mm]{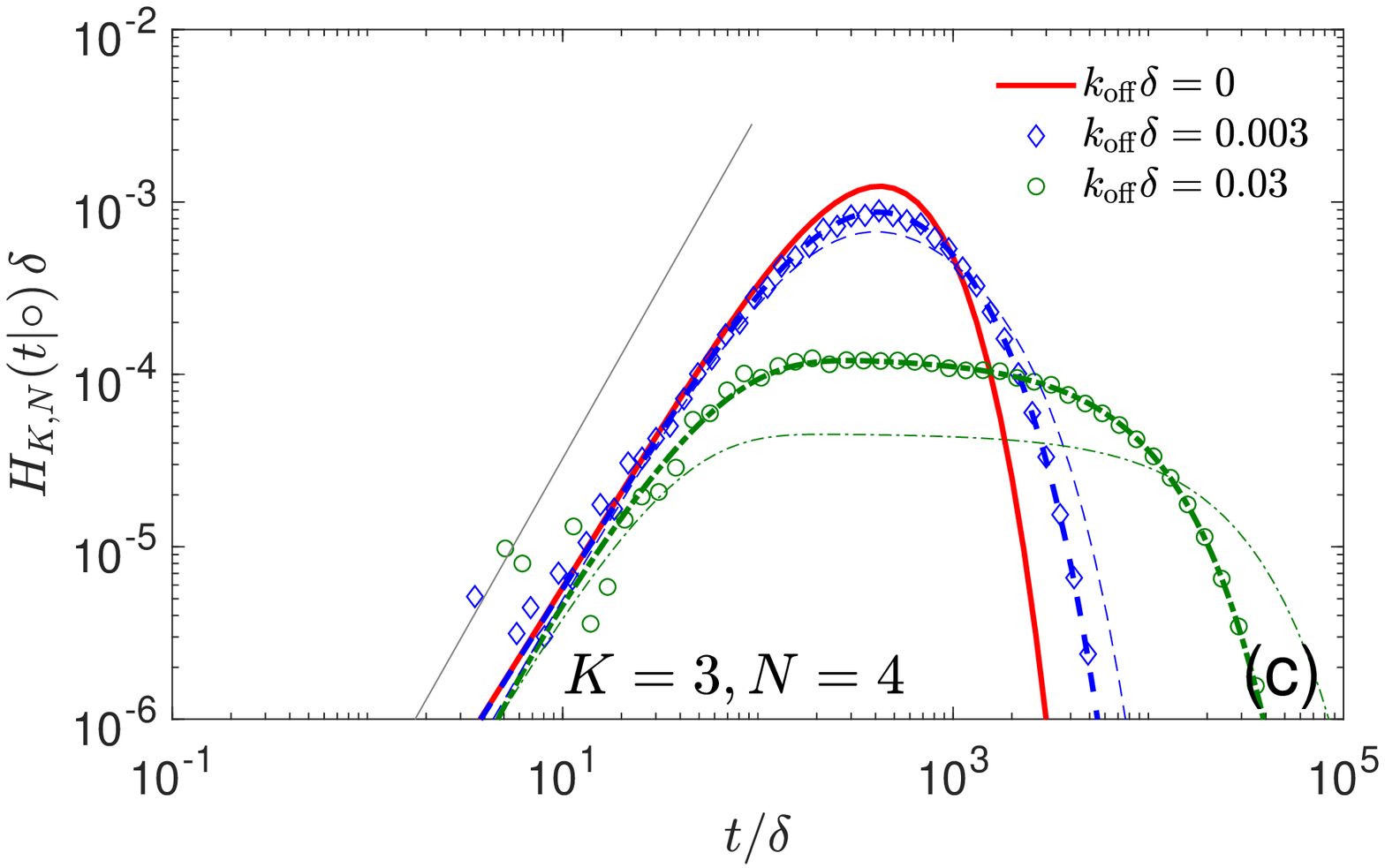}  
\includegraphics[width=77mm]{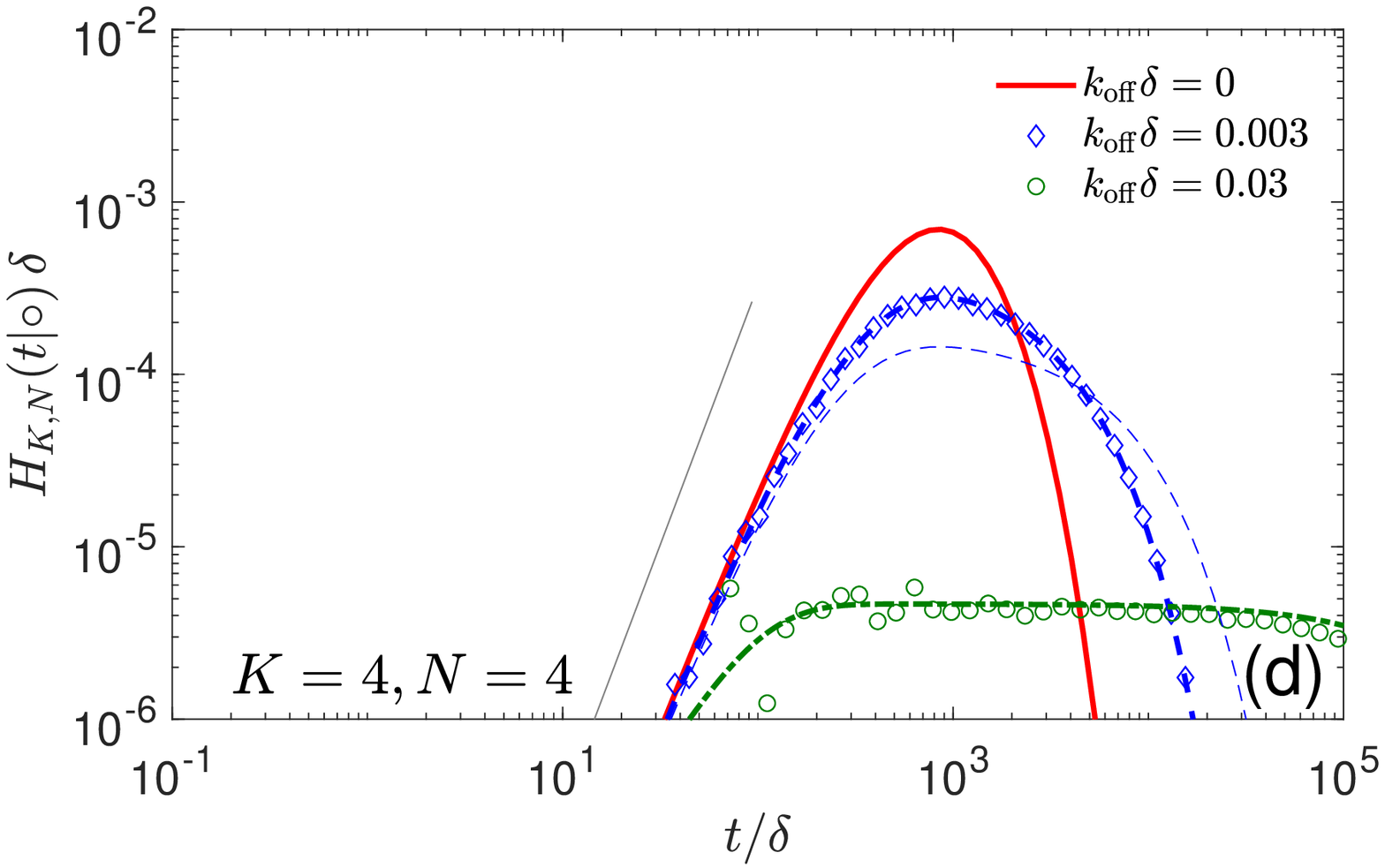}  
\end{center}
\caption{
Probability density of the reaction time $\T_{K,N}$ for restricted
diffusion between concentric spheres of radii $\rho$ and $R = 10
\rho$, with $N = 4$, $\kappa \rho/D = 1$, a timescale $\delta =
\rho^2/D$, three values of $\koff$ (see legend), and four values of
$K$: $K = 1$ {\bf (a)}, $K = 2$ {\bf (b)}, $K = 3$ {\bf (c)}, and $K =
4$ {\bf (d)}.  Symbols show empirical histograms from Monte Carlo
simulations with $10^6$ particles.  Thick solid line presents the
exact solution (\ref{eq:HKN_irrev}) for irreversible binding; thick
dashed lines indicate our approximation (\ref{eq:HKN_exact}) evaluated
numerically as described in \ref{sec:computation}.  Thin lines show
the Lawley-Madrid approximation (\ref{eq:HKN_LMA}), with $\nu$ given
by Eq. (\ref{eq:nu_LMA}); note that the thin line for the case
$\koff\delta = 0.03$ in panel {\bf (d)} is not visible as it appears
below the figure (i.e., $\bar{H}_{4,4}(t|\circ) \delta < 10^{-6}$).
Thin gray solid line presents the short-time asymptotic behavior
(\ref{eq:HKN_short}).}
\label{fig:H4}
\end{figure}

To illustrate our general results, we consider restricted diffusion
inside a confining reflecting sphere of radius $R$ towards a small
concentric partially reactive spherical target of radius $\rho$
(Fig. \ref{fig:scheme}(a)).  This domain can be considered as an
idealized model for the intracellular transport towards the nucleus or
a model of the presynaptic bouton \cite{Reva21}.  Figure
\ref{fig:H4} illustrates the behavior of the probability density
$H_{K,N}(t)$ for $N = 4$ and several values of $K$ in the case of a
small ($\rho/R = 0.1$), moderately reactive ($\kappa \rho/D = 1$)
target.  As the unbinding kinetics can only be initiated after the
first binding, the reaction time $\T_{1,N}$ is equal to the
first-binding time of the fastest particle and thus does not depend on
the unbinding rate $\koff$ (see also \ref{sec:cases}).  Expectedly,
three curves with different $\koff$ coincide on the panel
Fig. \ref{fig:H4}(a).  Moreover, the short-time behavior does not
depend on $\koff$ for any $K$.  In turn, the long-time decay is
strongly affected by $\koff$ when $K > 1$: the decay time $T_{K,N}$
increases with $\koff$ and thus the distribution is getting broader
for faster unbinding kinetics.  In all cases, the approximate solution
(\ref{eq:HKN_exact}) is in a remarkable agreement with Monte Carlo
simulations over a broad range of times.  We also stress that our
solution is exact for $K = N$.  The Lawley-Madrid approximation (see
\ref{sec:LMA}) captures correctly the overall behavior but
overestimates the decay time.  The agreement is better for smaller
$\koff$ and smaller $K$.  In turn, the disagreement for larger $\koff$
or $K$ is caused by moderate reactivity of the target, for which the
second condition in Eq. (\ref{eq:LMA_cond2}) is not satisfied.  Note
that the parameter $\epsilon$ from Eq. (\ref{eq:LMA_cond}) is equal to
$0.03$, wrongly suggesting the validity of the LMA.  This example
clearly illustrates that the single condition (\ref{eq:LMA_cond}) is
not sufficient and should be replaced by two separate conditions in
(\ref{eq:LMA_cond2}).  Figure \ref{fig:H4_kappa10} from
\ref{sec:other} illustrates that the disagreement is getting even
bigger for a small target with higher reactive $\kappa \rho/D = 10$.
In contrast, the LMA is very accurate for weakly reactive targets
(see, e.g., Fig. 4 in Ref. \cite{Lawley19}, which was plotted for the
case $\kappa \rho/D = 0.01$ and $\koff \rho^2/D = 0.001$).  Finally,
we emphasize that the short-time asymptotic relation
(\ref{eq:HKN_short}) is not accurate in the considered range of times,
requiring many correction terms for amendment (see
\ref{sec:asymptotics} for details).  Similar behavior was observed for
$N = 2$ and $N = 3$ (see Figs. \ref{fig:H2} and
\ref{fig:H3} from \ref{sec:other}).

\begin{figure}
\begin{center}
\includegraphics[width=77mm]{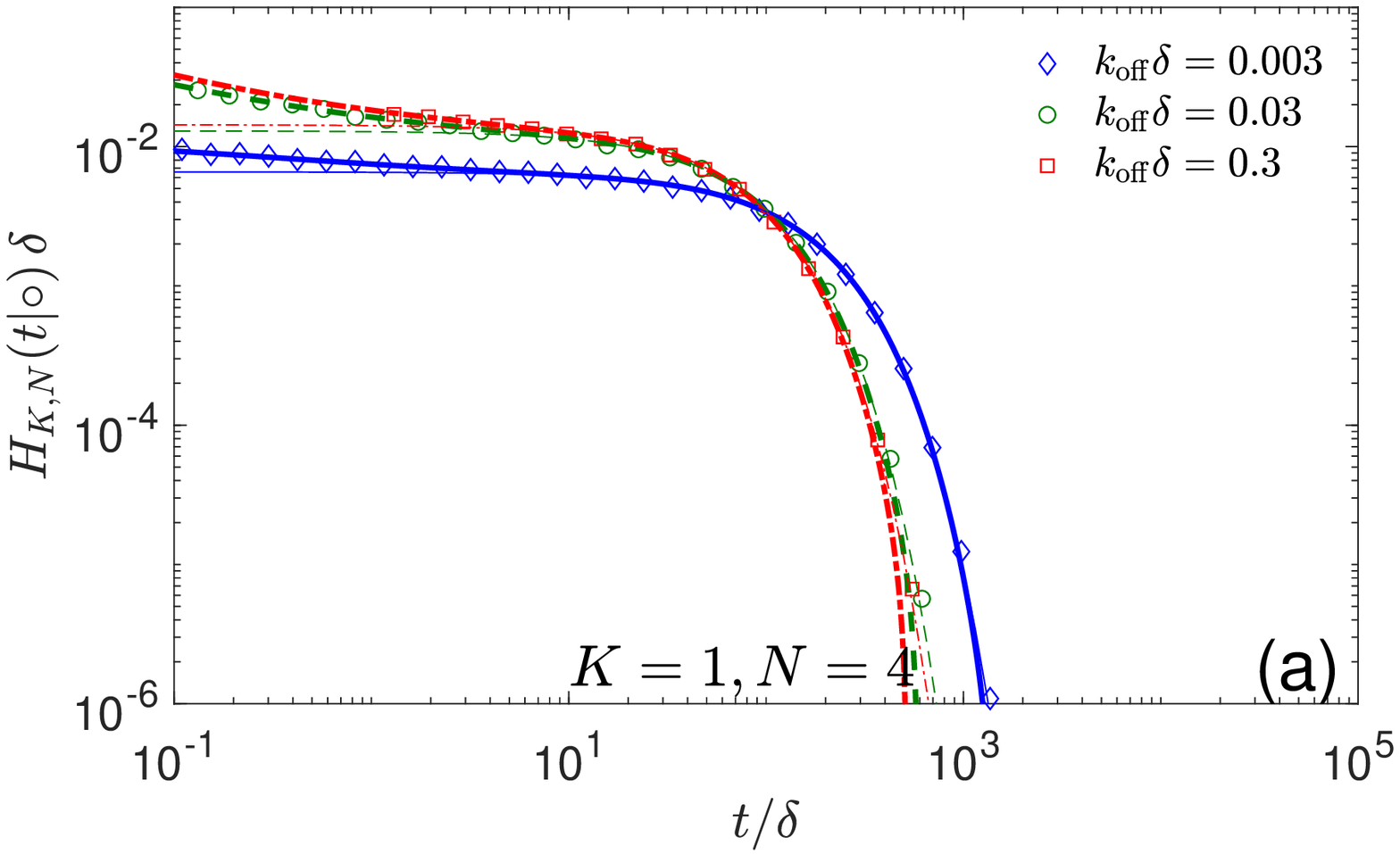} 
\includegraphics[width=77mm]{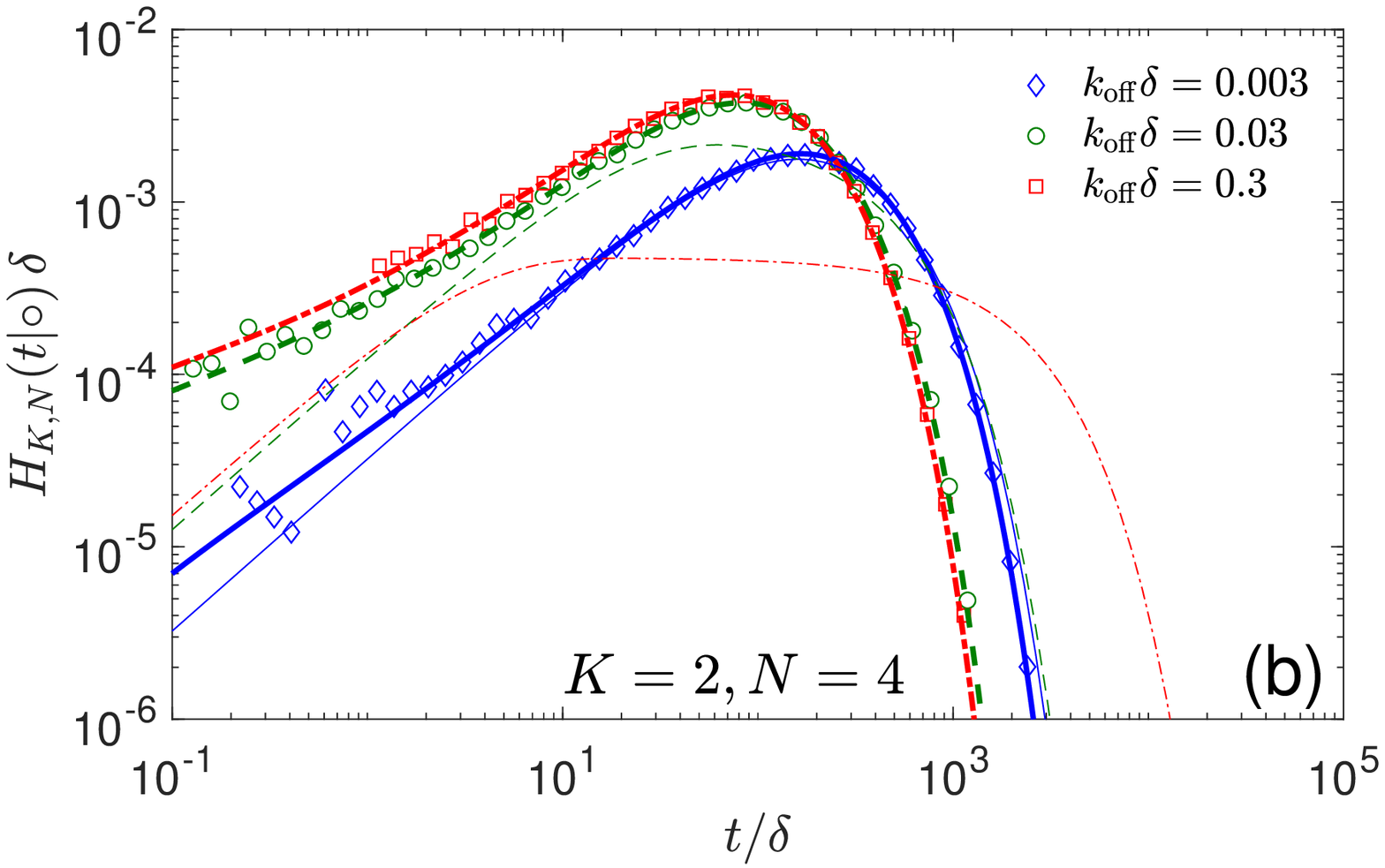} 
\includegraphics[width=77mm]{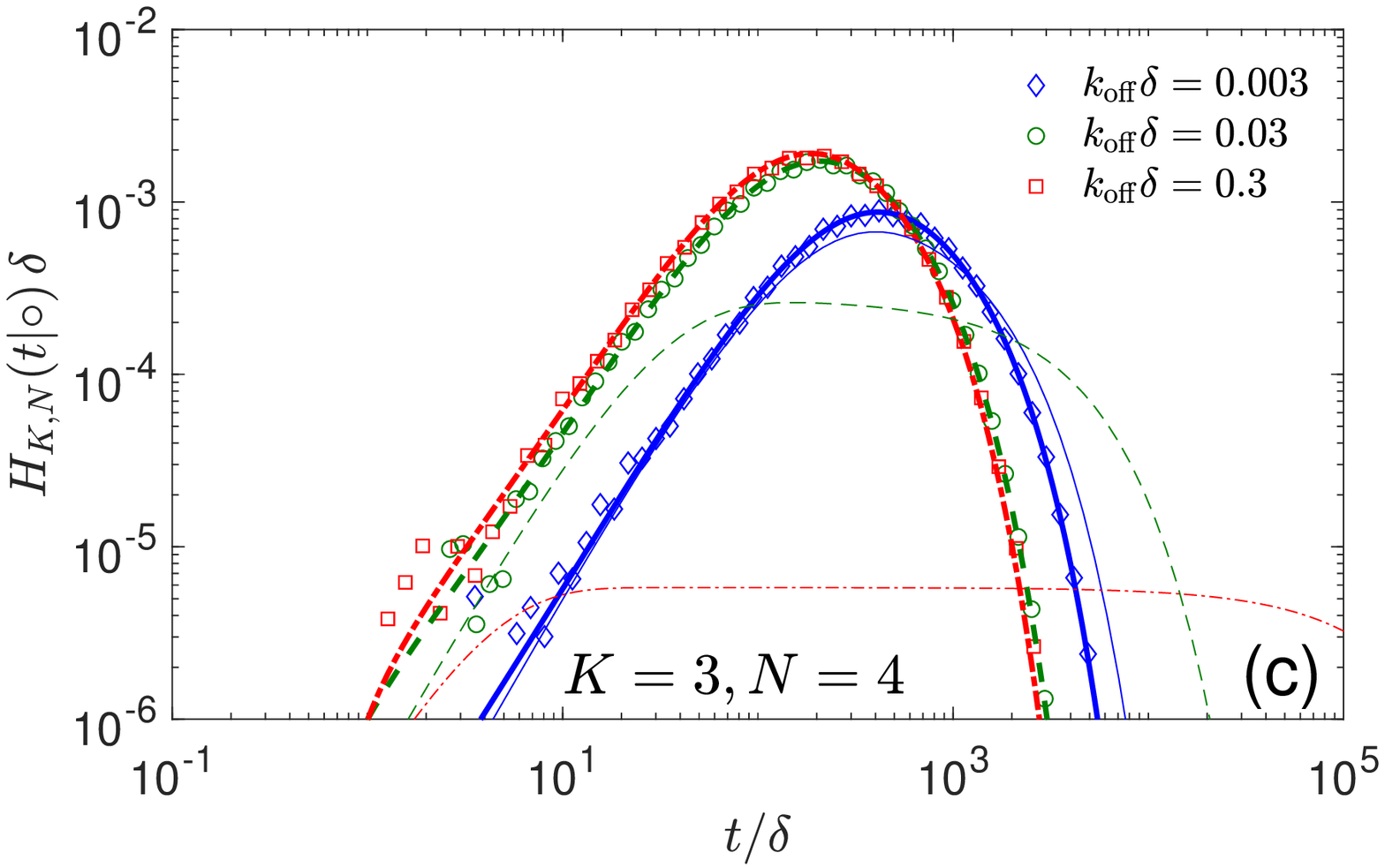} 
\includegraphics[width=77mm]{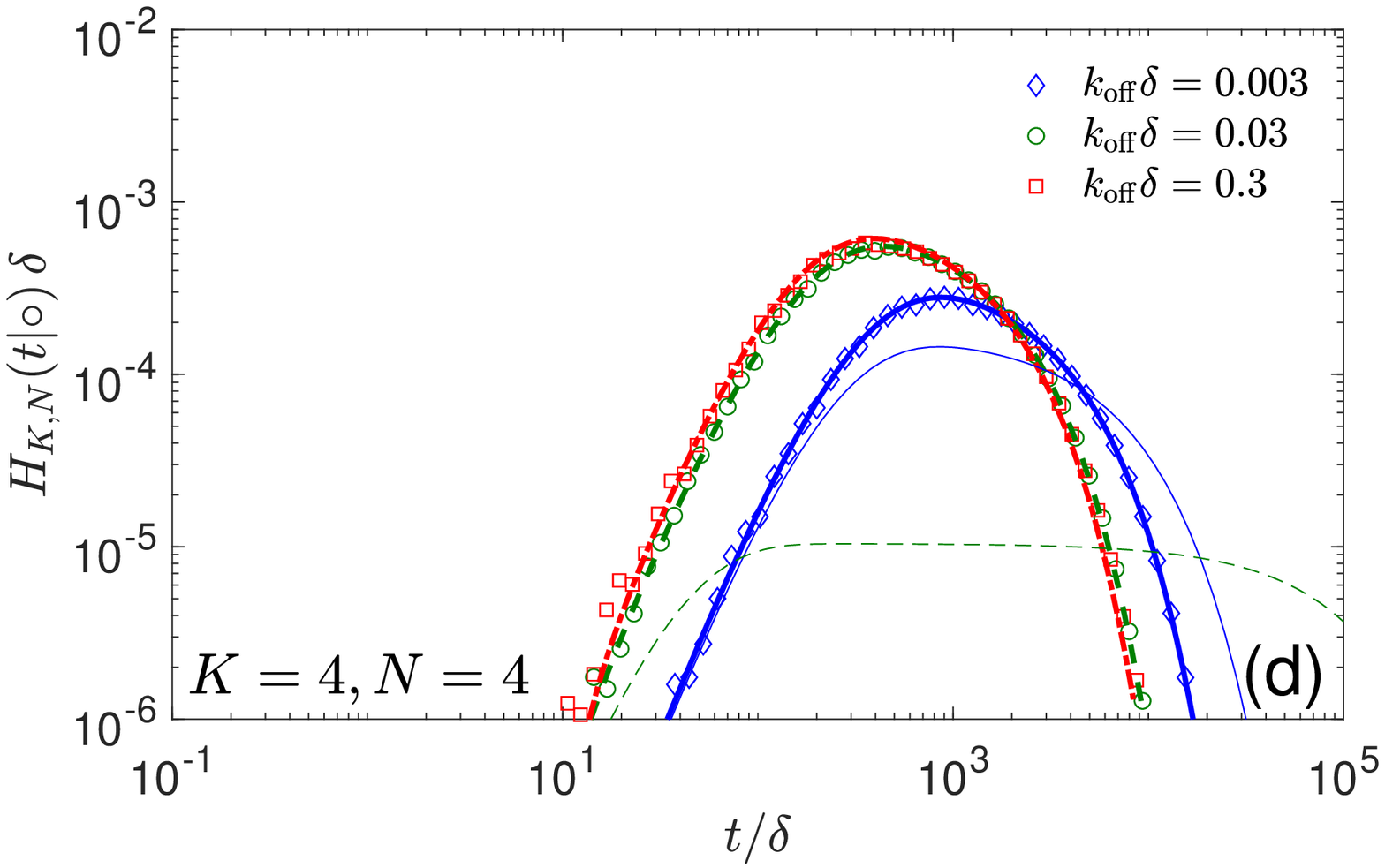} 
\end{center}
\caption{
Probability density of the reaction time $\T_{K,N}$ for restricted
diffusion between concentric spheres of radii $\rho$ and $R = 10
\rho$, with $N = 4$, a timescale $\delta = \rho^2/D$, three
combinations of $\koff$ and $\kappa$ ($\koff \delta = 0.003, 0.03,
0.3$ corresponding to $\kappa \rho/D = 1, 10, 100$, respectively, such
that $\eta = 1$ in all cases), and four values of $K$: $K = 1$ {\bf
(a)}, $K = 2$ {\bf (b)}, $K = 3$ {\bf (c)}, and $K = 4$ {\bf (d)}.
Symbols show empirical histograms from Monte Carlo simulations with
$10^6$ particles.  Thick lines indicate our approximation
(\ref{eq:HKN_exact}) evaluated numerically as described in
\ref{sec:computation}, whereas thin lines show the Lawley-Madrid
approximation (\ref{eq:HKN_LMA}), with $\nu$ given by
Eq. (\ref{eq:nu_LMA}); note that the thin line for the case
$\koff\delta = 0.03$ in panel {\bf (d)} is not visible as it appears
below the figure (i.e., $\bar{H}_{4,4}(t|\circ) \delta < 10^{-6}$).}
\label{fig:H4_comparison}
\end{figure}

The impact of unbinding kinetics and the consequent rebinding events
can be characterized by the dimensionless parameter $\eta = \koff
\langle \tau\rangle$, which is proportional to the ratio $\koff/\kon$
(or $\koff/\kappa$), see Eq. (\ref{eq:tau}).  In particular, this
parameter fully determines the steady-state probability
$P(\infty|\circ) = 1/(1+\eta)$ for a particle to be in the bound
state.  Intuitively, one might expect that $\eta$ mainly controls the
statistics of the reaction times $\T_{K,N}$.
To emphasize on the respective roles of binding and unbinding effects,
we fix $\eta = 1$ and compare the probability densities for three
combinations of $\kappa$ and $\koff$.  Figure \ref{fig:H4_comparison}
shows that two curves with larger unbinding rates $\koff (\rho^2/D) =
0.03$ and $\koff (\rho^2/D) = 0.3$ (and, accordingly, larger
reactivities) almost coincide.  This effect can be attributed to a
sort of statistical averaging due to multiple rebinding events.  In
contrast, the curve with the lowest $\koff$ and $\kappa$ differs from
the others, due to a limited number of rebinding events.  We conclude
that the parameter $\eta$ plays an important role but does not fully
determine the statistics of the reaction time.  Expectedly, the
Lawley-Madrid approximation gets less and less accurate as the
reactivity increases.

\begin{figure}
\begin{center}
\includegraphics[width=85mm]{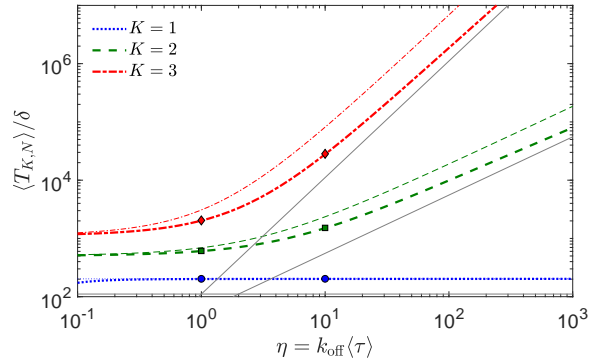} 
\end{center}
\caption{
Mean reaction time $\langle \T_{K,3}\rangle$ for restricted diffusion
between concentric spheres of radii $\rho$ and $R = 10 \rho$, with
$\kappa \rho/D = 1$, a timescale $\delta = \rho^2/D$, $N = 3$, and
three values of $K$ (see legend).  Thick lines show our approximation
(\ref{eq:Tmean}), thin lines present the Lawley-Madrid approximation
with $\nu$ given by Eq. (\ref{eq:nu_LMA}), while symbols illustrate
the results of Monte Carlo simulations with $10^6$ realizations.  Thin
straight solid lines present the large-$\eta$ asymptotic behavior
(\ref{eq:TKN_eta}).}
\label{fig:TK3}
\end{figure}

We complete this section by looking at the mean reaction time $\langle
\T_{K,N}\rangle$.  Figure \ref{fig:TK3} shows the dependence of
$\langle \T_{K,N}\rangle$ on the unbinding rate $\koff$ (rescaled by
$\langle\tau\rangle$) for a fixed reactivity $\kappa \rho/D = 1$.
When $\eta = \koff \langle\tau\rangle$ is small, the mean reaction
time is almost constant and close to $\langle \T_{K,N}^0\rangle$ for
irreversible binding ($\koff = 0$), as expected.  In turn, for $\eta
\gtrsim 1$, the mean reaction time starts to rapidly increase with
$\eta$.

\section{Conclusion}
\label{sec:conclusion}

In this paper, we investigated diffusion-controlled reactions or
events that are triggered on a target region after binding a
prescribed number $K$ among $N$ independently diffusing particles.
The reversible target-binding kinetics, which is so common for most
applications, presented the major mathematical difficulty.  We
developed a powerful theoretical approach to derive a new
approximation $H_{K,N}(t)$ for the probability density of the reaction
time $\T_{K,N}$ in the case when the particles were initially released
uniformly.  Under the assumption that the random positions of free
particles at time $\T_{K,N}$ remain to be uniform, we derived a
renewal equation that determines $H_{K,N}(t)$.  This convolution-type
equation was then solved in the Laplace domain to relate the
probability density via Eq. (\ref{eq:HKN_exact}) to two occupancy
probabilities, which were in turn expressed in terms of the survival
probability for a single particle.  In this way, we managed to
describe the collective effect of multiple diffusing particles in
terms of the diffusive dynamics of a single particle and thus to
extend the well-known extreme statistics for the $K$-th fastest
first-passage time to a more general and much more challenging setting
with reversible binding.  In other words, the knowledge of the
survival probability $S(t|\circ)$ (or, equivalently, $S(t)$) of a
single particle was sufficient for approximating the probability
density of the reaction time $\T_{K,N}$.

The assumption of uniform positions was the crucial step and the only
source of eventual deviations between the exact probability density
and our approximation (\ref{eq:HKN_exact}).  Strictly speaking, this
assumption is fulfilled exactly only for an inert non-reactive target
($\kappa = 0$).  When the target is reactive, binding events lead to a
formation of a depletion boundary layer near the target, in which the
probability density of finding a diffusing particle is lower, and thus
not uniform.  In contrast, unbinding events tend to homogenize the
probability density and thus render our assumption more accurate.  As
a consequence, our approximation is applicable whenever the
binding/unbinding kinetics ensure a nearly uniform distribution of
free particles.  A systematic study of quantitative conditions for the
validity of our approximation presents an important perspective of
this work in the future.  Meanwhile, Monte Carlo simulations that we
realized in this paper indicate that the approximation is remarkably
accurate when $\eta = \koff\langle\tau\rangle$ is not too small.  As
the limit $\eta = 0$ corresponds to irreversible binding (with either
$\koff = 0$, or $\kappa = \infty$), our approximation complements this
well-studied setting and thus provides the overall insight onto
diffusion-controlled reactions with multiple particles.

We also emphasize on the conceptual difference between our approach
and the Lawley-Madrid approximation.  The latter relied on the
exponential approximation for the survival probability of a single
particle, which is valid only for small and weakly reactive targets.
This restriction concerns only binding events and does not involve
unbinding kinetics.  In turn, our approximation deals with the exact
form of the survival probability, while the underlying assumption
depends on binding/unbinding kinetics.  As a consequence, it yields
accurate results even for highly reactive targets, if the unbinding
rate is not too small.  In summary, the validity range of our
approximation is different from that of the Lawley-Madrid
approximation (see details in \ref{sec:validity}), and it allows one
to deal with highly reactive targets.  At the same time, we outline
that the LMA is much more explicit and easier to implement and to
analyze, even in sophisticated geometric settings.  Moreover, the LMA
provides bounds to the first-crossing times for impatient particles.
These two approximations present therefore valuable and complementary
theoretical tools for studying diffusion-controlled reactions with
reversible target-binding kinetics.

The present work can be extended in several directions.  First, one
can further analyze and possibly relax the assumption of uniform
positions, beyond the discussion presented in \ref{sec:distribution}.
This analysis can potentially lead to an exact solution of the general
problem of impatient particles, which remains open for $1 < K < N$.
Second, one can consider more sophisticated diffusive dynamics such as
diffusing-diffusivity and switching models that allow one to
incorporate dynamic heterogeneities of the medium or reversible
binding to buffer molecules
\cite{Lanoiselee18,Sposini19,Grebenkov19e,Reva21}.  Similarly, more
elaborate target-binding mechanisms beyond that described by a
constant reactivity $\kappa$ can be investigated
\cite{Grebenkov20a,Grebenkov20f,Grebenkov22a,Grebenkov22b}.  For
instance, one can consider encounter-dependent reactivity that may
describe saturation effects after a number of reaction attempts that
are relevant to some chemical or biological reactions.  Moreover, one
can incorporate surface diffusion in the bound state that was shown to
enhance the overall reaction rate for a single particle
\cite{Benichou10,Benichou11,Rojo11,Rupprecht12a,Rupprecht12b,Rojo13,Benichou15}.
Finally, while the present paper focused on theoretical aspects of the
problem of impatient particles, its application to relevant examples
of diffusion-controlled events with multiple particles is a promising
perspective.  For this purpose, one needs further progress on the
numerical implementation of our approximation to deal with a large
number $N$ of diffusing particles (e.g., several hundred of calcium
ions).  A large-$N$ asymptotic analysis of the approximate solution
would also be beneficial.

\section*{Acknowledgements}
DG acknowledges the Alexander von Humboldt Foundation for support
within a Bessel Prize award.  AK was supported by the Prime Minister's
Research Fellowship (PMRF) of the Government of India.

%

\section*{Conflicts of interest}
There are no conflicts to declare.

\appendix

\section{Irreversible binding}
\label{sec:cases}

For irreversible binding ($\koff = 0$), the first-crossing time
$\T_{K,N}$ is identical to the $K$-th fastest first-passage time
$\Ti_{K,N}$ whose distribution is well known:
\begin{equation}
\P\{\Ti_{K,N} > t \} = \sum\limits_{j=0}^{K-1} {N \choose j} [S(t|\circ)]^{N-j} [1-S(t|\circ)]^{j} 
\end{equation}
and
\begin{equation}  \label{eq:HKN_irrev}
\fl
H^0_{K,N}(t) = - \frac{d\P\{\Ti_{K,N} > t \}}{dt} 
= K {N \choose K} [S(t|\circ)]^{N-K} [1-S(t|\circ)]^{K-1}  H(t|\circ),
\end{equation}
where $S(t|\circ)$ is the survival probability for a single particle
started uniformly, and $H(t|\circ) = -\frac{d}{dt} S(t|\circ)$ is the
probability density of the associated first-binding time (see
\ref{sec:distribution} and \ref{sec:computation} for details).

In the short-time limit, one can use the asymptotic relation
(\ref{eq:Hcirc_short}) for $H(t|\circ)$ to get
\begin{equation}   \label{eq:HKN_irrev_short}
H^0_{K,N}(t) \approx  \frac{K {N \choose K}}{\langle\tau\rangle^K}\, t^{K-1}  \qquad (t\to 0).
\end{equation}

In the case $K = 1$, the first-crossing time $\T_{1,N}$ for any
$\koff$ is equal to the first-passage time of the fastest particle,
$\T^0_{1,N}$, because unbinding kinetics does not matter here.  As a
consequence, one has the exact form:
\begin{equation}
\H_{1,N}(t) =  - \partial_t [S(t|\circ)]^N = N [S(t|\circ)]^{N-1} \, H(t|\circ).
\end{equation}

\section{Lawley-Madrid approximation}
\label{sec:LMA}

Lawley and Madrid developed an elegant approximate solution to the
general problem of impatient particles \cite{Lawley19}.  In the limit
of small and/or weakly reactive target such that
Eq. (\ref{eq:LMA_cond}) is fulfilled, the probability density of the
first-binding time for any starting point $\x_0$ was approximated by
an exponential density,
\begin{equation}  \label{eq:Ht_LMA}
H(t|\x_0) \approx \nu e^{-\nu t}, 
\end{equation}
with the rate $\nu$ determined by the smallest eigenvalue of the
Laplace operator.  In other words, the first-binding time $\tau_0$ and
the consequent rebinding times $\tau_k$ were assumed to be independent
exponential random variables.  Under this approximation, the number of
bound particles $\N(t)$ can be modeled by a Markovian birth-death
process $\bar{\N}(t)$ between $N+1$ states of $0,1,2,\ldots,N$ bound
particles:
\begin{equation}
\begin{tikzcd}[every arrow/.append style={shift left}]
 0 \arrow{r}{N\nu} & 1 \arrow{l}{\koff}  \arrow{r}{(N-1)\nu} & 2 \arrow{l}{2\koff}  
\quad \cdots \quad N-1 \arrow{r}{\nu} & N \arrow{l}{N\koff} 
\end{tikzcd}
\end{equation}
(bar denotes the quantities corresponding to the LMA).  Let $W$ be an
$(N+1)\times(N+1)$-dimensional matrix with zero elements except for
\begin{equation*}
W_{i,i+1} = i \koff,  \qquad W_{i+1,i} = (N+1-i) \nu \qquad (i=1,2,\ldots,N),
\end{equation*}
and $W_{i,i}$ are chosen so that $W$ has zero column sums.  The
distribution of the first-crossing time $\bar{\T}_{K,N} = \inf\{ t>0
~:~ \bar{\N}(t) = K\}$ can be written as \cite{Lawley19}
\begin{equation}  \label{eq:SKN_LMA}
\P\{ \bar{\T}_{K,N} > t\} = \sum\limits_{j=1}^K \left[\exp(W^{(K)} t)\right]_{j,1} \,,
\end{equation}
where $W^{(K)}$ is the $K\times K$ matrix obtained by retaining the
first $K$ columns and $K$ rows from $W$ and discarding everything
else, and the initial state was assumed to be $0$ (no bound
particle).  The probability density is
\begin{equation}  \label{eq:HKN_LMA}
\bar{H}_{K,N}(t) = \nu (N-K+1) \left[\exp(W^{(K)} t)\right]_{K,1} \,,
\end{equation}
while the mean time is fully explicit:
\begin{equation}  \label{eq:LMA_mean}
\langle \bar{\T}_{K,N} \rangle
= \frac{1}{\nu} \sum\limits_{m=1}^K \left(\frac{1}{b_m} + \sum\limits_{j=m+1}^K \frac{(\koff/\nu)^{j-m}}{b_j} 
\prod\limits_{i=m}^{j-1} \frac{d_i}{b_i}\right),
\end{equation}
with $b_m = N-K+m$ and $d_m = K-m$.  

In \cite{Grebenkov22}, we showed that in the case $K = N$, the LMA
captures qualitatively the behavior of the probability density
$\H_{N,N}(t)$.  However, it overestimates the mean reaction time and
the decay time, and totally fails at short times.  This is expected
because the LMA ignores the starting positions of the particles.

When the starting points of all particles are uniformly distributed,
the LMA turns out to be more accurate even at short times.  In fact,
the Taylor expansion of the exponential matrix in
Eq. (\ref{eq:HKN_LMA}) yields the correct power-law short-time
behavior:
\begin{eqnarray} \nonumber
\bar{H}_{K,N}(t) &\approx& \nu (N-K+1) \bigl[(W^{(K)})^{K-1}\bigr]_{K,1} t^{K-1} + O(t^K) \\
& = & \frac{N!}{(N-K)!} \, \nu^K  \, t^{K-1} + O(t^K), 
\end{eqnarray}
in which the lower-order terms were canceled due the tridiagonal
structure of the matrix $W^{(K)}$.  If $\nu$ was set to be
$1/\langle\tau \rangle$, the prefactor of this power law would differ
from the exact asymptotic relation (\ref{eq:HKN_short}) only by a
factor $\frac{1}{(K-1)!}$.  Moreover, the long-time behavior remains
qualitatively correct, even though the decay time is still
overestimated (see Figs. \ref{fig:H4} and \ref{fig:H4_comparison}).

\subsection*{Validity of the LMA}

Lawley and Madrid required the smallness of the parameter $\epsilon$
from Eq. (\ref{eq:LMA_cond}) for approximating the smallest eigenvalue
$\lambda_1$ of the Laplace operator in the confining domain $\Omega$
with mixed Robin-Neumann boundary condition on the boundary $\pa$ for
the associated eigenfunction $u_1(\x)$, 
\begin{equation*}
\left\{ \begin{array}{r l} D \partial_n u_1(\x) + \kappa\, u_1(\x) = 0 & (\x\in\Gamma) , \\
D \partial_n u_1(\x) = 0 & (\x\in\pa \backslash \Gamma) ,  \\  \end{array} \right.
\end{equation*}
where $\partial_n$ is the normal derivative oriented outwards the
domain $\Omega$.  Their approximation
\begin{equation}  \label{eq:lambda1}
\lambda_1 \approx \frac{\kappa |\Gamma|}{D|\Omega|}  \,.
\end{equation}
can be easily obtained by integrating the eigenvalue equation $-\Delta
u_1(\x) = \lambda_1 u_1(\x)$ over $\x\in\Omega$ and using the above
boundary condition:
\begin{equation}  \label{eq:lambda1_ratio}
\lambda_1 = - \frac{\int\nolimits_{\Gamma} d\x \, (\partial_n u_1(\x))}{\int\nolimits_{\Omega} d\x \, u_1(\x)}
= \frac{\kappa \int\nolimits_{\Gamma} d\x \, u_1(\x)}{D \int\nolimits_{\Omega} d\x \, u_1(\x)}  \,.
\end{equation}
The approximation (\ref{eq:lambda1}) follows immediately if $u_1(\x)$
is replaced by a constant.  This relation implies
\begin{equation}  \label{eq:nu_approx}
\nu = D\lambda_1 \approx \frac{\kappa |\Gamma|}{|\Omega|} = \frac{1}{\langle\tau\rangle} \,,
\end{equation}
in agreement with the fact that if the rebinding time $\tau$ is
assumed to obey an exponential law, its rate should be equal to the
inverse of the mean rebinding time.

However, the condition (\ref{eq:LMA_cond}) is not sufficient for
getting the approximation (\ref{eq:lambda1}).  For instance, in the
case of diffusion between concentric spheres with $\rho = 1$, $R =
10$, $D = 1$, and $\kappa = 1$, one has $\epsilon \approx 0.033$ and
$1/\langle\tau\rangle \approx 0.0030$, whereas the numerical solution
of Eq. (\ref{eq:lambda1_eq}), that determines the exact eigenvalue,
yields $D\lambda_1 \approx 0.0016$.  In other words, if one employs
the approximate relation (\ref{eq:nu_approx}) in this example, the
twofold error in the rate $\nu$ will be drastically amplified in the
computation of the mean reaction time $\langle \T_{K,N}\rangle$ or the
decay time $T_{K,N}$.  For this reason, Lawley and Madrid used the
numerically computed smallest eigenvalue for plotting their figures.

To further clarify this issue, it is instructive to analyze the
smallest eigenvalue $\lambda_1$.  For diffusion between concentric
spheres, the solution is summarized in \ref{sec:computation}.  In
particular, $\lambda_1$ is determined by the smallest strictly
positive solution of Eq. (\ref{eq:lambda1_eq}), whose asymptotic
behavior was given by Eq. (28) of Ref. \cite{Grebenkov18}.  When $\rho
\ll R$, a first-order approximation reads
\begin{equation}  \label{eq:lambda1_sphere}
\lambda_1 \approx \frac{\kappa |\Gamma|}{D|\Omega|(1 + \kappa \rho/D)} \,.
\end{equation}
In the case $\kappa \rho/D \ll 1$, we retrieve the approximate
relation (\ref{eq:lambda1}).  However, the smallness of the parameter
$\epsilon = \frac13 (\kappa \rho/D)(\rho/R)/(1+(\rho/R)^2)$ from
Eq. (\ref{eq:LMA_cond}) does not necessarily imply that $\kappa
\rho/D$ is small.  Actually, in the above example, we had $\kappa
\rho/D = 1$ that yielded the twofold smaller value of $\nu =
D\lambda_1$, as compared to $1/\langle \tau\rangle$.

An extension of Eq. (\ref{eq:lambda1_sphere}) to a general setting in
three dimensions was recently proposed in \cite{Chaigneau22}:
\begin{equation}  \label{eq:lambda1_domain}
\lambda_1 \approx \frac{\kappa |\Gamma|}{D|\Omega|(1 + \kappa |\Gamma|/(DC))} \,,
\end{equation}
where $C$ is the harmonic capacity (or capacitance) of the target
(e.g., $C = 4\pi \rho$ for a sphere of radius $\rho$).  This
approximation is valid when the target is small and located far away
from the outer reflecting boundary.  Qualitatively,
Eq. (\ref{eq:lambda1_domain}) can be interpreted as an interpolation
between two well-known limits: $\lambda_1 \approx C/|\Omega|$ for a
perfectly reactive target with $\kappa = \infty$
\cite{Mazya85,Ward93,Cheviakov11} and Eq. (\ref{eq:lambda1}) for an
almost inert target ($\kappa \to 0$).  One sees that the condition
\begin{equation}  \label{eq:LMA_cond3}
\kappa |\Gamma|/(DC) \ll 1
\end{equation}
ensures Eq. (\ref{eq:nu_approx}) and makes thus the exponential
approximation of the survival probability self-consistent.

We stress that the original derivation of the Lawley-Madrid
approximation in \cite{Lawley19} employed Eqs. (\ref{eq:Ht_LMA}) and
(\ref{eq:lambda1}) as distinct assumptions.  However, our relation
(\ref{eq:tau}) implies that these assumptions are actually tightly
related.  In fact, if the rebinding time is assumed to be
exponentially distributed according to Eq. (\ref{eq:Ht_LMA}), the rate
$\nu = D\lambda_1$ must be equal to the inverse of the mean rebinding
time $\langle\tau\rangle$, which in turn is equal to $|\Omega|/(\kappa
|\Gamma|)$ according to Eq. (\ref{eq:tau}).  As a consequence,
Eq. (\ref{eq:nu_approx}) can be considered as a necessary condition
for the applicability of the Lawley-Madrid approximation, which thus
requires that the target should be simultaneously small {\it and}
weakly reactive.

In order to ensure a proper comparison between our results and
the Lawley-Madrid approximation, we always set
\begin{equation}  \label{eq:nu_LMA}
\nu = D\lambda_1 = D \alpha_1^2/R^2,
\end{equation}
where $\alpha_1$ is the smallest strictly positive solution of
Eq. (\ref{eq:lambda1_eq}), which was obtained numerically.  In this
way, we tested directly the validity of a Markov birth-death process
representation of the system of impatient particles, which was the
cornerstone of the Lawley-Madrid approximation.  Note that setting
$\nu = 1/\langle\tau\rangle$ yielded worse results, which were not
shown in our figures.

\section{Distribution of a free particle}
\label{sec:distribution}

In this Appendix, we compute the probability density $P(\x,t|\x_0)$ of
finding a free particle that started from a point $\x_0$ at time $0$,
in the vicinity of a point $\x$ at time $t$.  For this purpose, we
extend the computation from Ref. \cite{Reva21,Grebenkov22} that
consists in adding up contributions according to the number of binding
events:
\begin{eqnarray*}
\fl
&& P(\x,t|\x_0) = G(\x,t|\x_0) 
+ \int\limits_0^t dt_1 \int\limits_{t_1}^t dt'_1 \, H(t_1|\x_0) \, \psi(t'_1-t_1) \, g(\x,t-t'_1) \\
\fl && 
+ \int\limits_0^t dt_1 \int\limits_{t_1}^t dt'_1 
\int\limits_{t'_1}^t dt_2 \int\limits_{t_2}^t dt'_2 \, H(t_1|\x_0) \, \psi(t'_1-t_1) 
H(t_2-t'_1) \, \psi(t'_2-t_2) \, g(\x,t-t'_2) + \ldots ,
\end{eqnarray*}
where $\psi(t) = \koff e^{-\koff t}$ is the probability density of the
waiting time on the target, and $H(t|\x_0)$ is the probability density
of the first-binding time for a particle started from $\x_0$.  The
first term represents the contribution without binding, with
$G(\x,t|\x_0)$ being the propagator for a single particle in the
presence of a partially reactive target.  The second term includes the
contribution with a single binding at time $t_1$, staying on the
target up to time $t'_1$, at which the particle unbinds and resumes
its diffusion to $\x$, where
\begin{equation}
g(\x,t) = \frac{1}{|\Gamma|} \int\limits_{\Gamma} d\x_0 \, G(\x,t|\x_0)
\end{equation}
is the propagator for a particle that started from a uniformly
distributed point on the target $\Gamma$.  The third term counts two
bindings events: binding at $t_1$, unbinding at $t'_1$, binding at
$t_2$, unbinding at $t'_2$, and arrival in $\x$ at $t$, where
\begin{equation}  \label{eq:Ht}
H(t) = \frac{1}{|\Gamma|} \int\limits_{\Gamma} d\x_0 \, H(t|\x_0)
\end{equation} 
is the probability density of the rebinding time (given that the
unbound particle is released from a uniformly distributed point on the
target).  The fourth, fifth and next terms correspond to $3$, $4$,
$\ldots$ binding events.  In the Laplace domain, one gets
\begin{eqnarray*}  \nonumber
\tilde{P}(\x,p|\x_0) &=& \tilde{G}(\x,p|\x_0) + \tilde{H}(p|\x_0) \, \frac{\koff}{p + \koff} \, \tilde{g}(\x,p) \\
&+& \tilde{H}(p|\x_0) \, \frac{\koff}{p + \koff} \,\tilde{H}(p) \, \frac{\koff}{p + \koff}\,  \tilde{g}(\x,p) + \ldots \\  
&=& \tilde{G}(\x,p|\x_0) + \tilde{H}(p|\x_0) \, \frac{\koff}{p + \koff(1 - \tilde{H}(p))} \, \tilde{g}(\x,p) ,
\end{eqnarray*}
where all terms were summed up as a geometric series, and tilde
denotes Laplace transformed quantities, e.g.,
\begin{equation*}
\tilde{f}(p) = \L\{ f(t)\}(p) = \int\limits_0^\infty dt \, e^{-pt} \, f(t).
\end{equation*}
Since the probability density $H(t|\x_0)$ can be understood as the
integral of the probability flux density over the target region, one
gets
\begin{eqnarray*}
H(t|\x_0) &=& \int\limits_{\Gamma} d\x \, (-D\partial_n G(\x,t|\x_0))  
= \int\limits_{\Gamma} d\x \, (\kappa G(\x,t|\x_0)) = \kappa |\Gamma| \, g(\x_0,t),
\end{eqnarray*}
i.e.,
\begin{equation}
\tilde{g}(\x,p) = \frac{\langle \tau \rangle}{|\Omega|} \tilde{H}(p|\x) ,
\end{equation}
where we used Eq. (\ref{eq:tau}) for the mean rebinding time
$\langle\tau\rangle$, and the Robin boundary condition on the target
region.  We conclude that
\begin{equation}  \label{eq:Ptilde}
\tilde{P}(\x,p|\x_0) = \tilde{G}(\x,p|\x_0) +  \frac{\tilde{H}(p|\x_0) \,\koff \langle\tau\rangle \, \tilde{H}(p|\x)}
{|\Omega|(p + \koff(1 - \tilde{H}(p)))} .
\end{equation}

Similarly, if $P_0(\x,t)$ denotes the probability density for a
particle that was initially bound to the target, to be in the vicinity
of a point $\x$ at time $t$, one gets in the Laplace domain:
\begin{eqnarray*}
\tilde{P}_0(\x,p) &=& \tilde{\psi}(p) \,\tilde{g}(\x,p) + \tilde{\psi}(p)\, \tilde{H}(p)\, \tilde{\psi}(p)\,  \tilde{g}(\x,p) + \ldots 
= \frac{\tilde{\psi}(p) \tilde{g}(\x,p)}{1 - \tilde{H}(p) \,\tilde{\psi}(p)}  \,,
\end{eqnarray*}
that yields
\begin{equation}  \label{eq:P0}
P_0(\x,t) = \frac{\koff \langle\tau\rangle}{|\Omega|} \, P(t|\x) ,
\end{equation}
where $P(t|\x)$ is the occupancy probability of the target (see
also \ref{sec:single}).

\subsection*{Normalization}

It is instructive to check that the probability density $P(\x,t|\x_0)$
is correctly normalized.  For this purpose, we recall that the Green's
function $\tilde{G}(\x,p|\x_0)$ satisfies the boundary value problem
\begin{equation}
\left\{ \begin{array}{ll} 
(p - D\Delta_{\x}) \tilde{G}(\x,p|\x_0) = \delta(\x-\x_0)  & (\x\in\Omega), \\
(D\partial_{n} + \kappa \I_{\Gamma}(\x))\tilde{G}(\x,p|\x_0) = 0 & (\x\in\pa), \\ \end{array} \right.
\end{equation}
where $\Delta_{\x}$ is the Laplace operator acting on $\x$,
$\delta(\x-\x_0)$ is the Dirac distribution, and $\I_{\Gamma}(\x)$ is
the indicator function of $\Gamma$: $\I_{\Gamma}(\x) = 1$ for
$\x\in\Gamma$, and $0$ otherwise.  The second relation is the mixed
Robin-Neumann boundary condition representing reflections on the inert
boundary $\pa\backslash\Gamma$, and partial reactivity on the target
region $\Gamma$.
The integral of the first relation over $\x\in\Omega$ yields
\begin{equation}  \label{eq:Gp_circ}
\int\limits_{\Omega} d\x \, \tilde{G}(\x,p|\x_0) = \tilde{S}(p|\x_0) = \frac{1 - \tilde{H}(p|\x_0)}{p} \,,
\end{equation}
where $\tilde{S}(p|\x_0)$ is the Laplace-transformed survival
probability.  Similarly, as $\tilde{H}(p|\x_0)$ satisfies
\begin{equation}
\left\{ \begin{array}{ll} 
(p - D\Delta_{\x_0}) \tilde{H}(p|\x_0) = 0  & (\x_0\in\Omega), \\
(D\partial_{n} + \kappa \I_{\Gamma}(\x_0))\tilde{H}(p|\x_0) = \kappa \I_{\Gamma}(\x_0) & (\x_0\in\pa), \\ \end{array} \right.
\end{equation}
the integral of the first relation over $\x_0\in\Omega$ yields
\begin{equation}  
\int\limits_{\Omega} d\x_0 \, \tilde{H}(p|\x_0) = \kappa |\Gamma| \frac{1 - \tilde{H}(p)}{p} ,
\end{equation}
where we used the Green's formula and the above boundary condition for
$\tilde{H}(p|\x_0)$, while $\tilde{H}(p)$ is the Laplace transform of
$H(t)$ defined by Eq. (\ref{eq:Ht}).  In the limit $p\to 0$, the
left-hand side approaches $|\Omega|$ due the normalization of
$H(t|\x_0)$, whereas the right-hand side goes to $\kappa |\Gamma|
\langle\tau\rangle$, from which Eq. (\ref{eq:tau}) for the mean
rebinding time $\langle \tau\rangle$ follows.  We get thus
\begin{equation}  \label{eq:Hp_circ}
\tilde{H}(p|\circ) \equiv \frac{1}{|\Omega|} \int\limits_{\Omega} d\x_0 \, \tilde{H}(p|\x_0) = \frac{1 - \tilde{H}(p)}{p\langle\tau\rangle}
= \frac{\tilde{S}(p)}{\langle\tau\rangle} \,,
\end{equation}
where $\circ$ denotes the average over uniformly distributed starting
point.  This relation implies that
\begin{equation}  \label{eq:Hcirc_S}
H(t|\circ) = \frac{S(t)}{\langle\tau \rangle} 
\end{equation}
is a monotonously decreasing function of time.  Note also that the
Taylor expansion of Eq. (\ref{eq:Hp_circ}) allows one to express the
moments of the first-binding time $\tau_\circ$, e.g.,
\begin{equation}  \label{eq:tau_circ_mean}
\langle\tau_{\circ}\rangle = \int\limits_0^\infty dt \, t\, H(t|\circ) = \frac{\langle\tau^2\rangle}{2\langle\tau\rangle} \,.
\end{equation}
We outline that $\tau_\circ$ is the first-binding time for a particle
that started uniformly in the bulk $\Omega$, whereas $\tau$ is the
rebinding time (i.e., the first-binding time for a particle that
started uniformly on the target).  Combining Eqs. (\ref{eq:Gp_circ},
\ref{eq:Hp_circ}), the integral of Eq. (\ref{eq:Ptilde}) over
$\x\in\Omega$ reads
\begin{equation}  \label{eq:Ptilde_int}
\int\limits_{\Omega} d\x \, \tilde{P}(\x,p|\x_0) = \frac{1}{p} - \frac{\tilde{H}(p|\x_0)}{p + \koff(1-\tilde{H}(p))} \,,
\end{equation}
where the last term is the Laplace transform of the occupancy
probability $P(t|\x_0)$ of the target for a particle that started from
$\x_0$, see also Eq. (\ref{eq:Px0}).  Moving the last term to the
left-hand side, one sees that the normalization is indeed satisfied:
\begin{equation}  \label{eq:Px_normalization}
P(t|\x_0) + \int\limits_{\Omega} d\x \, P(\x,t|\x_0) = 1 .
\end{equation}

Similarly, the integral of Eq. (\ref{eq:P0}) reads in the Laplace
domain:
\begin{eqnarray*}  
\int\limits_\Omega d\x \, \tilde{P}_0(\x,p) &=& \koff \langle\tau\rangle \, \tilde{P}(p|\circ) = \koff \tilde{Q}(p) \tilde{S}(p)  \\
&=& \tilde{Q}(p)\bigl( \koff \tilde{S}(p) + 1\bigr) - \tilde{Q}(p) = \frac{1}{p} - \tilde{Q}(p) ,
\end{eqnarray*}
where $\tilde{Q}(p)$ is the Laplace transform of the occupancy
probability $Q(t)$ of the target for a particle that was initially
bound, see also Eq. (\ref{eq:Qp}).  The above relation implies the
normalization of $P_0(\x,t)$:
\begin{equation}  \label{eq:P0_normalization}
Q(t) + \int\limits_\Omega d\x \, P_0(\x,t) = 1.
\end{equation}

\subsection*{Long-time behavior}

In the long-time limit, $G(\x,t|\x_0)$ vanishes exponentially fast and
does not contribute.  In turn, the second term in
Eq. (\ref{eq:Ptilde}) yields as $p\to 0$:
\begin{equation}
\frac{\tilde{H}(p|\x_0) \,\koff \langle\tau\rangle \, \tilde{H}(p|\x)}
{|\Omega|(p + \koff(1 - \tilde{H}(p)))} \approx \frac{\koff \langle\tau\rangle}{|\Omega| p(1 + \koff \langle \tau\rangle)} \,, 
\end{equation}
so that
\begin{equation}
\lim\limits_{t\to\infty} P(\x,t|\x_0) = \frac{1 - P_\infty}{|\Omega|} \,,
\end{equation}
where
\begin{equation}  \label{eq:Pinfty}
P_\infty = \frac{1}{1 + \koff \langle \tau\rangle} \,.
\end{equation}
In other word, unbinding events ensure that the position of a free
particle in the long-time limit is distributed uniformly inside the
domain, as expected.

\subsection*{Uniformly distributed starting points}

When all particles start initially from uniformly distributed points,
one defines
\begin{eqnarray}  \nonumber
\tilde{P}(\x,p|\circ) &\equiv& \frac{1}{|\Omega|} \int\limits_{\Omega} d\x_0 \, \tilde{P}(\x,p|\x_0) 
= \frac{1}{p|\Omega|} - \frac{\tilde{H}(p|\x)}{|\Omega|(p + \koff (1-\tilde{H}(p)))} \,,
\end{eqnarray}
where we used Eq. (\ref{eq:Ptilde_int}) and the symmetry $P(\x,t|\x_0)
= P(\x_0,t|\x)$.  In the time domain, we get thus
\begin{equation}   \label{eq:Pxcirc}
P(\x,t|\circ) = \frac{1 - P(t|\x)}{|\Omega|} \,.
\end{equation}

Why $P(\x,t|\circ)$ is not uniform?  At short times, the main
contribution to the probability density of arriving at $\x$ comes from
the trajectories started close to that point.  If $\x$ is far from the
target, the probability of binding the target $P(t|\x)$ is very small,
and thus $P(\x,t|\circ)$ is almost constant.  In turn, if $\x$ is
close to the target, the particles started from its neighborhood have
higher chances to bind to the target and thus be in the bound state at
time $t$.  As a consequence, $P(\x,t|\circ)$ is smaller near the
target; this is similar to the formation of a depletion zone near a
reactive target.  The difference is that, as time goes on, all
particles, irrespective of their starting points, start to experience
the same effect of reversible binding, and $P(\x,t|\circ)$ is getting
uniform (in contrast to the case of a reactive target with
irreversible binding when the depletion zone would grow and finally
exhaust all particles).

\section{Occupancy probabilities}
\label{sec:single}

In Ref. \cite{Grebenkov22}, the focus was on the case when the
particles start from a fixed point $\x_0$ and search for a partially
reactive target $\Gamma$ with reactivity $\kappa$, from which they can
unbind at rate $\koff$.  The statistics of the first-crossing time
$\T_{N,N}$ was determined by two occupancy probabilities: the
probability $Q(t)$ of finding the particle in the bound state at time
$t$ given that it was bound at time $0$, and the probability
$P(t|\x_0)$ of finding the particle in the bound state at time $t$
given that it was initially released from a point $\x_0$.  Both
probabilities were found explicitly in the Laplace domain in the same
way as presented in \ref{sec:distribution}:
\begin{equation}  \label{eq:Qp}
\tilde{Q}(p) = \frac{1}{p + \koff(1 - \tilde{H}(p))}
\end{equation}
and
\begin{equation}  \label{eq:Px0}
\tilde{P}(p|\x_0) = \tilde{H}(p|\x_0) \, \tilde{Q}(p) ,
\end{equation}
where $\tilde{H}(p)$ is the Laplace transform of the probability
density of the rebinding time $\tau$, see Eq. (\ref{eq:Ht}).

If the starting point $\x_0$ is uniformly distributed, $P(t|\x_0)$
should be replaced by 
\begin{equation}  \label{eq:Pcirc_def}
P(t|\circ) \equiv \frac{1}{|\Omega|} \int\limits_{\Omega} d\x_0 \, P(t|\x_0),
\end{equation}
where $\circ$ indicates the uniform starting point.  According to
Eqs. (\ref{eq:P0}, \ref{eq:P0_normalization}), one gets
Eq. (\ref{eq:Pcirc}).  One sees that $\tilde{Q}(p)$ and thus
$\tilde{P}(p|\circ)$ are expressed in terms of $\tilde{H}(p)$.  Note
also that Eqs. (\ref{eq:Hp_circ}, \ref{eq:Px0}, \ref{eq:Pcirc_def})
yield
\begin{equation}  \label{eq:Ptilde_circ}
\tilde{P}(p|\circ) = \tilde{Q}(p) \tilde{H}(p|\circ) \,,
\end{equation}
which in the time domain reads
\begin{equation}  \label{eq:Pt_circ0}
P(t|\circ) = \int\limits_0^t dt' Q(t') H(t-t'|\circ)  .
\end{equation}

Alternatively, if $\{p_n\}$ are the poles of $\tilde{P}(p|\x_0)$, the
residue theorem allows one to invert the Laplace transform to get (if
all poles are simple):
\begin{equation}
P(t|\x_0) = P_\infty + \sum\limits_{n=1}^\infty v_n(\x_0) \, e^{p_n t}  ,
\end{equation}
where $v_n(\x_0)$ is the residue of $\tilde{P}(p|\x_0)$ at the pole
$p_n$, and $P_\infty$ is the residue at pole $p_0 = 0$ (that we treat
separately, see \cite{Grebenkov22} for details).  As a consequence,
\begin{equation}  \label{eq:Pcirc0}
P(t|\circ) = P_\infty + \sum\limits_{n=1}^\infty \hat{v}_n \, e^{p_n t} ,
\end{equation}
where
\begin{equation}   \label{eq:vhat}
\hat{v}_n = \frac{1}{|\Omega|} \int\limits_{\Omega} d\x \, v_n(\x) ,
\end{equation}
from which
\begin{equation}   \label{eq:Q}
Q(t) = P_\infty - \eta \sum\limits_{n=1}^\infty \hat{v}_n \, e^{p_n t} ,
\end{equation}
with $\eta = \koff \langle\tau\rangle$.

\subsection*{Short-time asymptotic behavior}

In the short-time limit, the target region can be considered as
locally flat so that $\tilde{H}(p|\x_0)$ can be approximated by
$\tilde{H}_{\rm hl}(p|\delta) = e^{-\delta \sqrt{p/D}}/(1 +
\sqrt{pD}/\kappa)$ for the half-line, where $\delta$ is the distance
to the boundary.  As a consequence, $\tilde{H}(p)
\approx 1/(1+\sqrt{pD}/\kappa)$ and thus 
\begin{equation}
\tilde{H}(p|\circ) \approx \frac{1}{p\langle\tau\rangle(1 + \kappa/\sqrt{pD})} \approx \frac{1 - \kappa/\sqrt{pD}}{p\langle\tau\rangle} + O(p^{-2}) , 
\end{equation}
from which
\begin{equation}  \label{eq:Hcirc_short}
H(t|\circ) \approx \frac{1}{\langle\tau\rangle} \biggl(1 - \frac{2\kappa \sqrt{Dt}}{\sqrt{\pi} D} + O(t)\biggr) \qquad (t\to 0),
\end{equation}
and thus 
\begin{equation}
1-S(t|\circ) \approx \frac{t}{\langle\tau\rangle} \biggl(1 - \frac{4\kappa \sqrt{Dt}}{3\sqrt{\pi} D} + O(t)\biggr)
\end{equation}
and
\begin{equation}
P(t|\circ) \approx \frac{t}{\langle\tau\rangle} + O(t^{3/2}) \qquad (t\to 0).
\end{equation}
Note also that Eq. (\ref{eq:Pcirc}) implies a monotonous decrease of
$P(t|\circ)$ with time: $dP(t|\circ)/dt \geq 0$.  In addition,
Eqs. (\ref{eq:Hcirc_S}, \ref{eq:Hcirc_short}) imply that
\begin{eqnarray}
S(t) &\approx& 1 - \frac{2\kappa \sqrt{Dt}}{\sqrt{\pi} D} + O(t)   \qquad (t\to 0), \\
H(t) &\approx& \frac{\kappa }{\sqrt{\pi} \sqrt{Dt}} + O(1)   \qquad (t\to 0).
\end{eqnarray}

\section{Numerical computation}
\label{sec:computation}

\subsection*{Probability density}

Following \cite{Grebenkov22}, we integrate by parts the convolution
(\ref{eq:HKN_exact}) to transform it into an integral equation
\begin{equation}  \label{eq:integral}
\overline{\Pr_t(K|K)} - \Pr_t(K|0) = S_{K,N}(t) - \int\limits_0^t dt \, S_{K,N}(t-t') \biggl(-\partial_{t'} \overline{\Pr_{t'}(K|K)}\biggr) ,
\end{equation}
where $S_{K,N}(t) = \P\{ \T_{K,N} > t\}$ is the approximate survival
probability, and we used that $S_{K,N}(0) = \overline{\Pr_0(K|K)} =
1$.  Here $\Pr_t(K|0)$ and $\overline{\Pr_t(K|K)}$ are expressed via
Eqs. (\ref{eq:PrK0}, \ref{eq:PrKK}) in terms of $P(t|\circ)$ and
$Q(t)$, which in turn are given by Eqs. (\ref{eq:Pcirc0}, \ref{eq:Q}).
For diffusion between concentric spheres, the poles $\{p_n\}$ and the
related residues were determined in Ref. \cite{Reva21,Grebenkov22}.
Note that the integral of the function $v_n(\x)$ in
Eq. (\ref{eq:vhat}) can be found explicitly.  After discretization of
the integral in Eq. (\ref{eq:integral}) over a linear grid, we
evaluate $S_{K,N}(t)$ and then $H_{K,N}(t)$ by applying the fast
Fourier transform to resolve the convolution problem (see details in
Ref. \cite{Grebenkov22}).

\subsection*{Monte Carlo simulations}

For Monte Carlo simulations, we use a standard event-driven scheme
described in detail in Ref. \cite{Grebenkov22}.  The only difference
concerns the generation of the first-binding times that are governed
by the probability density $H(t|\circ)$ instead of $H(t|\x_0)$.  This
probability density and the related survival probability $S(t|\circ)$
can be found from their spectral expansions:
\begin{eqnarray}  \label{eq:So_spectral}
S(t|\circ) &=& \sum\limits_{n=1}^\infty a_n \, e^{-Dt\lambda_n}  ,\\
H(t|\circ) &=& D \sum\limits_{n=1}^\infty \lambda_n \, a_n \, e^{-Dt\lambda_n}  ,
\end{eqnarray}
where $\lambda_n$ are the eigenvalues of the Laplace operator in
$\Omega$, and
\begin{equation}  \label{eq:an}
a_n = \frac{1}{|\Omega|} \left|\int\limits_\Omega d\x \, u_n(\x)\right|^2
\end{equation}
are the coefficients obtained from the $L_2$-normalized eigenfunctions
$u_n(\x)$.  As their computation is detailed in
Ref. \cite{Grebenkov22}, we only recall that the eigenvalues are
determined as $\lambda_n = \alpha_n^2/R^2$, where $\{\alpha_n\}$ are
strictly positive solutions of the trigonometric equation
\cite{Lawley19}:
\begin{equation}  \label{eq:lambda1_eq}
- \frac{\alpha^2+1}{1 - \alpha\, \ctan((1-\rho/R) \alpha)} - \frac{R}{\rho} + 1 = \frac{\kappa R}{D}  \,,
\end{equation}
which is equivalent to Eq. (B9) from Ref. \cite{Grebenkov22}.  In
turn, the coefficients $a_n$ are
\begin{eqnarray*}
\fl
a_n &=& 6\rho^4 \mu \frac{(R-\rho)\alpha_n \cos(\alpha_n \beta) - (\rho+R\alpha_n^2)\sin(\alpha_n\beta)}{\alpha_n^3(R^3-\rho^3) } \\
\fl
&\times&  \biggl( \bigl(\mu \rho^2 - R(R-\rho)\alpha_n^2\bigr) \cos(\alpha_n\beta) 
- \bigl(R(R-\rho)\mu + (R^2+\rho^2)\bigr) \alpha_n\sin(\alpha_n\beta)  \biggr)^{-1} ,
\end{eqnarray*}
where $\mu = \kappa\rho/D$ and $\beta = R/\rho - 1$.

A generated array of independent random realizations of the reaction
times $\T_{K,N}$ is used to compute the mean value, $\langle
\T_{K,N}\rangle$, and the empirical probability density of $\T_{K,N}$.
As the probability density $\H_{K,N}(t)$ typically spans several orders
of magnitude in time, we produce a renormalized histogram $h(z)$ of
$\zeta = \ln \T_{K,N}$ and then draw $h(z)/e^{z}$ versus $t = e^{z}$,
see Figs. \ref{fig:H4} and \ref{fig:H4_comparison}.

\section{Asymptotic behavior}
\label{sec:asymptotics}

\subsection*{Short-time limit}

In the short-time limit, unbinding kinetics does not matter so that
$H_{K,N}(t) \approx H^0_{K,N}(t)$, where $H^0_{K,N}(t)$ is given by
Eq. (\ref{eq:HKN_irrev}) and its short-time asymptotic behavior
(\ref{eq:HKN_irrev_short}) implies Eq. (\ref{eq:HKN_short}).  However,
Fig. \ref{fig:H4} shows a considerable deviation from this behavior
because it is achieved only at very short times, at which the
probability density is too small and thus not relevant.

In order to clarify this point, we focus on diffusion between
concentric spheres and compute next-order terms of the probability
density $H(t|\circ)$ as $t\to 0$.  For this purpose, we analyze the
large-$p$ behavior of its Laplace transform,
\begin{equation}  \label{eq:Hp_circ_sphere}
\tilde{H}(p|\circ) = \frac{\frac{3 \rho D}{p(R^3-\rho^3)}\bigl((R-\rho)\alpha + (\rho R\alpha^2-1) \tanh \xi\bigr) }
{R\alpha - \tanh \xi + \frac{D}{\kappa \rho}\, (\xi + (\rho R \alpha^2 -1)\tanh \xi)} \,,
\end{equation}
where $\alpha = \sqrt{p/D}$ and $\xi = \alpha (R-\rho)$.  In the limit
$p\to\infty$, $\tanh(\xi)$ can be replaced by $1$, with exponentially
small corrections:
\begin{equation*}
\tilde{H}(p|\circ) \approx \frac{1}{p\langle\tau\rangle}  - 
\frac{\frac{\kappa \rho}{\alpha^2 \, D^2\langle\tau\rangle} (R\alpha-1)}{\alpha(R-\rho) 
+ \rho R\alpha^2 - 1 + \frac{\kappa \rho}{D}(R\alpha-1)} \,.
\end{equation*}
This expression can be decomposed into partial fractions as
\begin{eqnarray*}
\fl
\tilde{H}(p|\circ) &\approx& \frac{1}{p\langle\tau\rangle}  - \frac{\kappa \rho}{D^2\langle\tau\rangle} 
\left( \frac{1}{(1+\mu)\alpha^2} - \frac{\rho}{(1+\mu)^2\alpha} 
+ \frac{\rho^2}{(1+\mu)^2(\alpha \rho + 1+\mu)} \right) ,
\end{eqnarray*}
where $\mu = \kappa \rho/D$.  The inverse Laplace transform yields
\begin{eqnarray*}
\fl
H(t|\circ) &\approx& \frac{1}{\langle\tau\rangle(1 + \kappa \rho/D)}  
+ \frac{\kappa \rho^2}{\sqrt{\pi} D\langle\tau\rangle(1+\mu)^2} 
\, \frac{1 - \sqrt{\pi}\, E_{\frac12,\frac12}\bigl(-(1+\mu)\sqrt{Dt}/\rho\bigr)}{ \sqrt{Dt}}  \,,
\end{eqnarray*}
where $E_{\alpha,\beta}(z)$ is the Mittag-Leffler function:
\begin{equation}
E_{\alpha,\beta}(z) = \sum\limits_{n=0}^\infty \frac{z^n}{\Gamma(\alpha n+ \beta)} 
\end{equation}
(here the Euler function $\Gamma(z)$ should not be confused with our
notation $\Gamma$ for the target region).  Using the identity
$E_{\alpha,\beta}(z) = z E_{\alpha,\alpha+\beta}(z) +
1/\Gamma(\beta)$, we get
\begin{equation}   \label{eq:Ho_short}
H(t|\circ) \approx \frac{1}{\langle\tau\rangle(1 + \kappa \rho/D)} 
\biggl(1 + \frac{\kappa \rho}{D} \, E_{\frac12,1}\bigl(-(1/\rho +\kappa/D)\sqrt{Dt}\bigr) \biggr)  \,.
\end{equation}  
The short-time expansion reads then
\begin{equation} 
H(t|\circ) \approx \frac{1}{\langle\tau\rangle} \left(1 + \frac{\mu}{1+\mu} 
\sum\limits_{n=1}^\infty \frac{(-(1+\mu)\sqrt{Dt}/\rho)^n}{\Gamma(\frac12 n+1)} \right).
\end{equation}
This expansion can be truncated to few terms when
$(1+\mu)\sqrt{Dt}/\rho \ll 1$.  However, when this condition is not
satisfied, one needs many terms to get an accurate result.  This is
precisely what happens in Fig. \ref{fig:H4}, in which the short-time
behavior is established for $t/\delta = Dt/\rho^2 \sim 10$, at which
the above condition is not fulfilled.  In this case, it is more
convenient to keep the Mittag-Leffler function (note also that
$E_{\frac12,1}(-z) = \erfcx(z) = e^{z^2} \erfc(z)$ is the scaled
complementary error function).  However, Eq. (\ref{eq:Ho_short}) is
specific to the case of concentric spheres and is not applicable for
general domains.

From Eq. (\ref{eq:Ho_short}), we can also obtain the short-time
behavior of the survival probability:
\begin{eqnarray*}
1 - S(t|\circ) & = & \int\limits_0^t dt' \, H(t'|\circ) \approx \frac{1}{\langle\tau\rangle(1+\mu)}  
 \biggl( t + \mu t  E_{\frac12,2}\bigl(-(1+\mu)\sqrt{Dt}/\rho\bigr) \biggr) ,
\end{eqnarray*}
where we used the identity:
\begin{equation}
\int\limits_0^z dz' \, E_{\alpha,1}(z^\alpha) = z \, E_{\alpha,2}(z^\alpha) .
\end{equation}
Using the identity,
\begin{equation}
E_{\frac12,2}(-c\sqrt{t}) = 1 - \frac{4c\sqrt{t}}{3\sqrt{\pi}} + c^2 t E_{\frac12,1}(-c\sqrt{t}) ,
\end{equation}
one also gets
\begin{eqnarray*}
1 - S(t|\circ) &\approx& \frac{t}{\langle\tau\rangle} \left( 1 - \frac{4\mu \sqrt{Dt}}{3\sqrt{\pi} \rho} 
+ \frac{\mu(1+\mu) Dt}{\rho^2}  E_{\frac12,1}\bigl(-(1+\mu)\sqrt{Dt}/\rho\bigr) \right) .
\end{eqnarray*}

\subsection*{Long-time limit}

At long times, the probability density $H_{K,N}(t)$ decays
exponentially according to Eq. (\ref{eq:HKN_long}), with the decay
time $T_{K,N}$ determined by the largest (negative) pole $p_c$ of
$\tilde{H}_{K,N}(p)$, which is given by the largest (negative) zero of
$\L\{\overline{\Pr_t(K|K)}\}(p)$.  Following the approach from
\cite{Grebenkov22}, we get
\begin{equation}
p_c \approx \Pr_\infty(K|K) \left(\int\limits_0^\infty dt \bigl(\overline{\Pr_t(K|K)} - \overline{\Pr_\infty(K|K)}\bigr)\right)^{-1} ,
\end{equation}
from which the decay time $T_{K,N}$ can be approximated by
Eq. (\ref{eq:TKN_decay}).

\section{Mean reaction time}
\label{sec:mean}

\subsection*{Derivation}

In this Appendix, we derive and analyze an approximation for the mean
reaction time:
\begin{eqnarray*}
\langle \T_{K,N}\rangle &=&  - \lim\limits_{p\to 0} \frac{\partial \tilde{\H}_{K,N}(p|\circ)}{\partial p} 
\approx  - \lim\limits_{p\to 0} \frac{\partial \tilde{H}_{K,N}(p|\circ)}{\partial p} \\
&=& \lim\limits_{p\to 0} \left(\frac{\L\{ t\Pr_t(K|0)\}}{\L\{\overline{\Pr_{t}(K|K)}\}} 
-  \frac{\L\{\Pr_t(K|0)\} \, \L\{t \overline{\Pr_{t}(K|K)}\} }{(\L\{\overline{\Pr_{t}(K|K)}\})^2} \right),
\end{eqnarray*}
where we used our approximation (\ref{eq:HKN_exact}).  Setting
\begin{eqnarray}
a_k &=& \int\limits_0^\infty dt \, t^k \biggl(\Pr_t(K|0) - \Pr_\infty(K|0)\biggr), \\
b_k &=& \int\limits_0^\infty dt \, t^k \biggl(\Pr_{t}(K|K) - \Pr_{\infty}(K|K)\biggr) ,
\end{eqnarray}
one can employ Taylor expansions of the above Laplace transforms to
get
\begin{equation}
\langle \T_{K,N}\rangle \approx \frac{\Pr_\infty(K|0) b_0}{[\overline{\Pr_{\infty}(K|K)}]^2} - \frac{a_0}{\overline{\Pr_{\infty}(K|K)}} \,.
\end{equation}
Using the identity
\begin{equation}
\sum\limits_{j=0}^K {K \choose j} {N-K \choose j} = {N \choose K} ,
\end{equation}
one can check that
\begin{equation}  \label{eq:Pr_infty}
\Pr_\infty(K|0) = \overline{\Pr_{\infty}(K|K)} = {N \choose K} P_\infty^K (1-P_\infty)^{N-K} \,,
\end{equation}
so that
\begin{equation}
\langle \T_{K,N}\rangle \approx \frac{b_0 - a_0}{\Pr_\infty(K|0)}   \,,
\end{equation}
which can be rewritten in a more explicit form as
Eq. (\ref{eq:Tmean}).  The same technique can be used to get
higher-order moments.  We emphasize that this relation is not
applicable for irreversible binding because $\koff = 0$ implies
$P_\infty = 1$ and thus $\Pr_\infty(K|0) = 0$ for any $K < N$.  In
turn, for $K = N$, Eq. (\ref{eq:Tmean}) remains valid even for $\koff
= 0$ and coincides with the exact relation derived in Ref.
\cite{Grebenkov22}.

\begin{figure}
\begin{center}
\includegraphics[width=85mm]{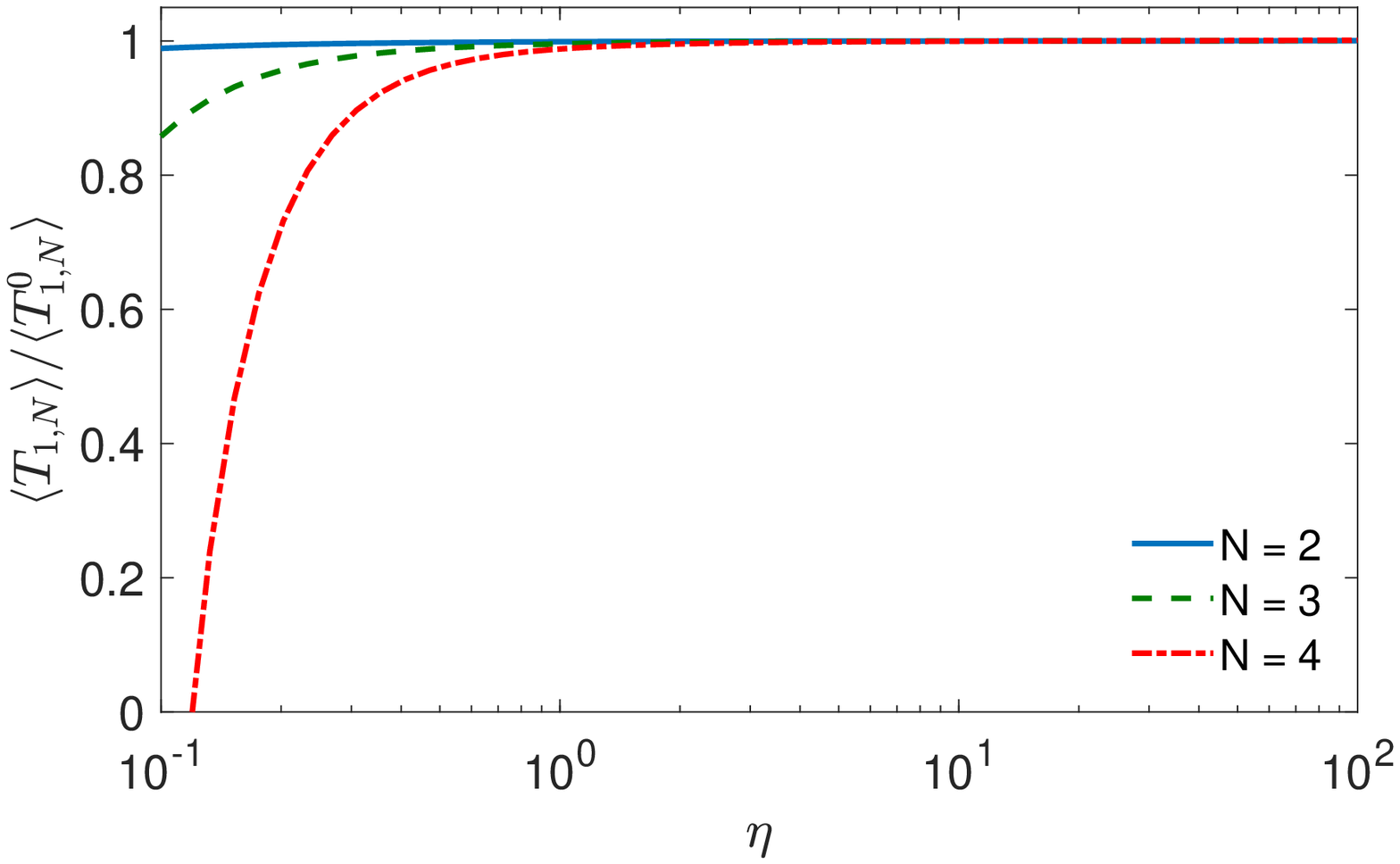} 
\end{center}
\caption{
The ratio between the approximation (\ref{eq:Tmean}) of the mean
reaction time $\langle \T_{1,N}\rangle$ and the exact form
(\ref{eq:Tmean_irrev}) of the mean reaction time $\langle
\T_{1,N}^0\rangle$ for irreversible binding, as a function of $\eta =
\koff\langle\tau\rangle$, for restricted diffusion between concentric
spheres of radii $\rho$ and $R = 10 \rho$, with $\kappa \rho/D = 1$,
and three values of $N$ (see legend). }
\label{fig:T1N}
\end{figure}

\subsection*{Validity}

We stress that the above derivation is based on the approximate
relation (\ref{eq:HKN_exact}) so that Eq. (\ref{eq:Tmean}) is an
approximation of the mean reaction time.  We recall that our
approximation relied on the assumption that the $N-K$ free particles
are uniformly distributed at the time when the threshold crossing
event happens.  According to Eq. (\ref{eq:Pcirc}), this assumption is
better fulfilled when 
\begin{equation}   \label{eq:Pcirc_bound}
P(t|\circ) \leq P_\infty =  \frac{1}{1+\eta} \ll 1, 
\end{equation}
i.e., when $\eta = \koff\langle\tau\rangle$ is large.  In contrast,
when $\koff \to 0$, unbinding events are rare and thus do not allow to
spread away the depletion zone near the target.  As a consequence, our
assumption is not applicable, and the derived approximate formulas may
fail.  Note that in the limit $\koff = 0$, the mean reaction time is
given by
\begin{equation}  \label{eq:Tmean_irrev}
\langle \T_{K,N}^0 \rangle = \int\limits_0^\infty dt \, t \, \H_{K,N}^0(t) ,
\end{equation}
with $\H_{K,N}^0(t)$ being determined by the exact relation
(\ref{eq:HKN_irrev}).

The failure of our approximation can be illustrated by taking the
limit $\koff\to 0$, for which the numerator of Eq. (\ref{eq:Tmean})
should vanish, yielding an identity
\begin{equation}  \label{eq:identity_S}
\int\limits_0^\infty dt [S(t|\circ)]^{N-K} \biggl(1 - {N\choose K} [1-S(t|\circ)]^K\biggr) = 0
\end{equation}
for any $K < N$.  This identity is satisfied for $S(t|\circ) = e^{-\nu
t}$, i.e., if the first-binding time obeys an exponential distribution
with a rate $\nu$.  We note that this is also related to the
assumption of the Lawley-Madrid approximation, see further discussion
in \ref{sec:validity}.  We emphasize that the identity
(\ref{eq:identity_S}) does not hold in general, thus invalidating
Eq. (\ref{eq:Tmean}) in the limit $\koff
\to 0$.

Figure \ref{fig:T1N} illustrates the validity range of the approximate
relation (\ref{eq:Tmean}).  Here we plot the ratio between the
approximate value of $\langle \T_{1,N}\rangle$ from Eq.
(\ref{eq:Tmean}), and the exact value $\langle \T_{1,N}^0 \rangle$
from Eq. (\ref{eq:Tmean_irrev}).  As binding of the first particle
does not depend on the unbinding kinetics, this ratio should be equal
to $1$ for any $\koff$.  In turn, deviations from $1$ highlight
limitations of the approximate relation (\ref{eq:Tmean}).  For $N =
2$, the ratio remains close to $1$ for the considered range of $\eta =
\koff\langle\tau\rangle$.  As $N$ increases, one observes deviations
from $1$ for $\eta \lesssim 1$.  A more systematic study is needed for
establishing quantitative criteria of the validity range of the
developed approximation.

\subsection*{Asymptotic behavior}

When $\eta$ is large enough, the inequality (\ref{eq:Pcirc_bound})
implies $\Pr_\infty(K|0) \approx {N \choose K} P_\infty^K$ and
$\overline{\Pr_t(K|K)} \approx [Q(t)]^K$, from which
\begin{equation}
\langle\T_{K,N}\rangle \approx \frac{1}{{N \choose K} P_\infty^K} \int\limits_0^\infty dt \biggl([Q(t)]^K - P_{\infty}^K\biggr).
\end{equation}
In this regime, the mean reaction time $\langle\T_{K,N}\rangle$ is
close to the mean reaction time $\langle\T_{K,K}\rangle$ divided by
the combinatorial factor ${N\choose K}$.  The latter was investigated
in Ref. \cite{Grebenkov22}, and it was shown to behave as $(1 +
\eta)^K/(\koff K)$ for large $K$.  Neglecting $1$ in comparison to
$\eta \gg 1$, one deduces Eq. (\ref{eq:TKN_eta}).  Strictly speaking,
this relation is valid for $N \geq K \gg 1$ but Fig. \ref{fig:TK3}
suggests that this asymptotic relation can be used for any $K > 1$ if
$\eta$ is large enough.

Note that in the case $K = 1$, one can compute the integral exactly by
using the small-$p$ asymptotic behavior of $\tilde{Q}(p)$:
\begin{equation}
\langle \T_{1,1}\rangle \approx \frac{1}{P_\infty} \int\limits_0^\infty dt (Q(t) - P_\infty) 
= \frac{\koff \langle\tau^2\rangle}{2(1+\koff \langle\tau\rangle)} \,.
\end{equation}
As $\koff \to 0$, this express vanishes, indicating again the failure
of our approximation.  In turn, as $\koff\to\infty$, one gets the
limit $\langle\tau^2\rangle/(2\langle\tau\rangle) =
\langle\tau_\circ\rangle$ according to Eq. (\ref{eq:tau_circ_mean}).
In other words, we retrieve the exact value of the mean first-passage
time $\langle \T_{1,1}\rangle = \langle \T_{1,1}^0\rangle =
\langle\tau_\circ\rangle$ for $N = 1$.

\section{Other illustrations}
\label{sec:other}

\begin{figure}
\begin{center}
\includegraphics[width=77mm]{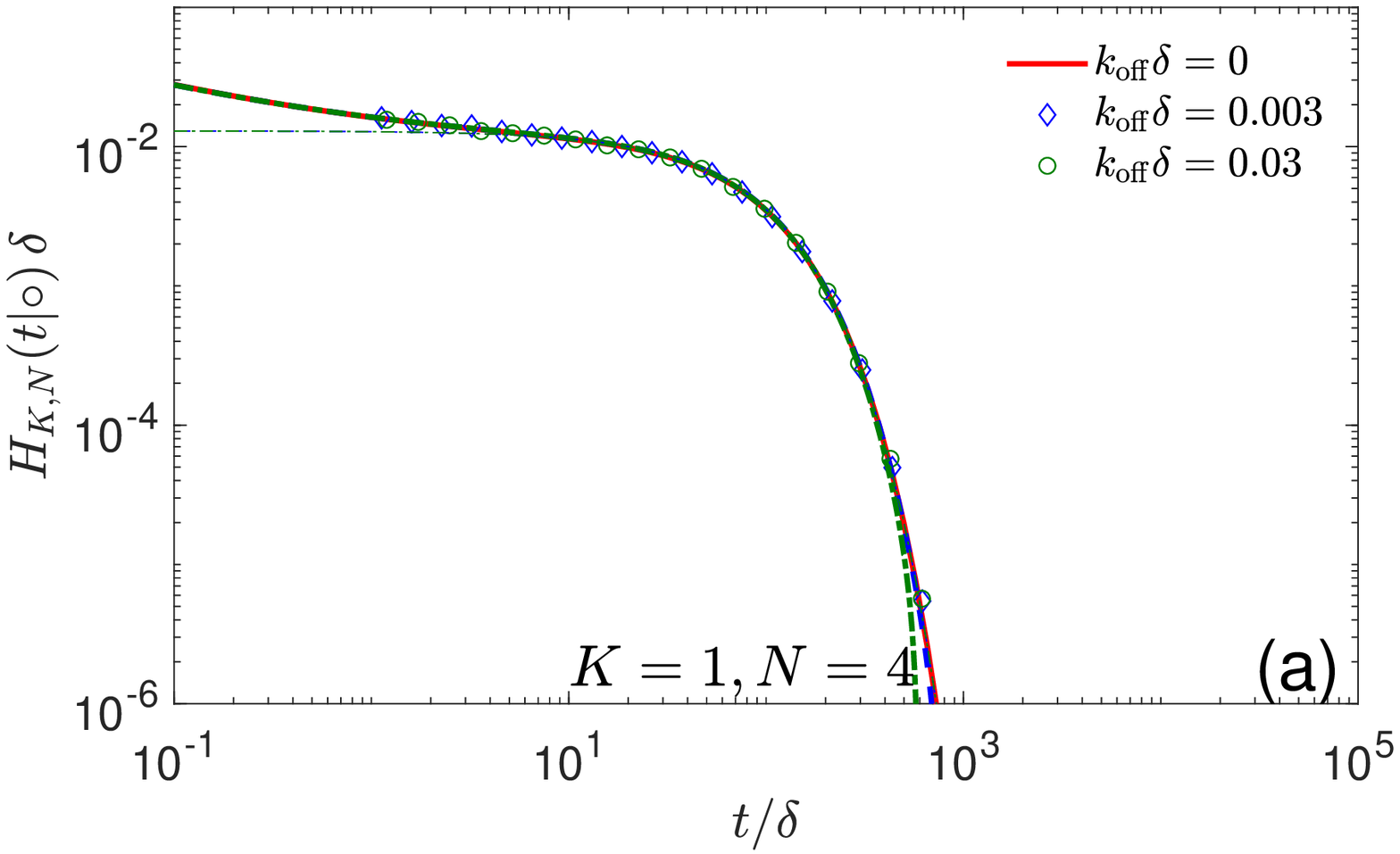} 
\includegraphics[width=77mm]{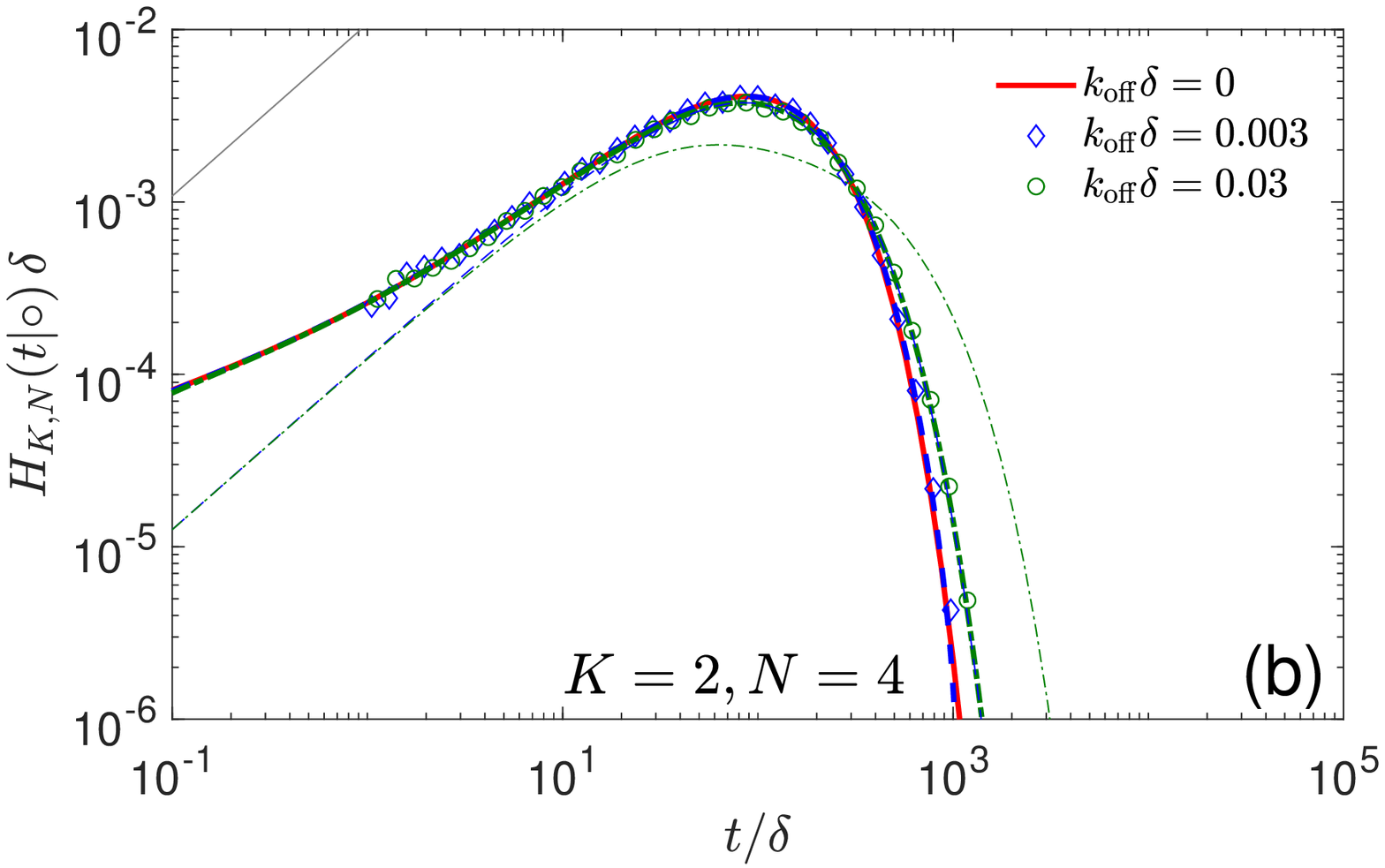} 
\includegraphics[width=77mm]{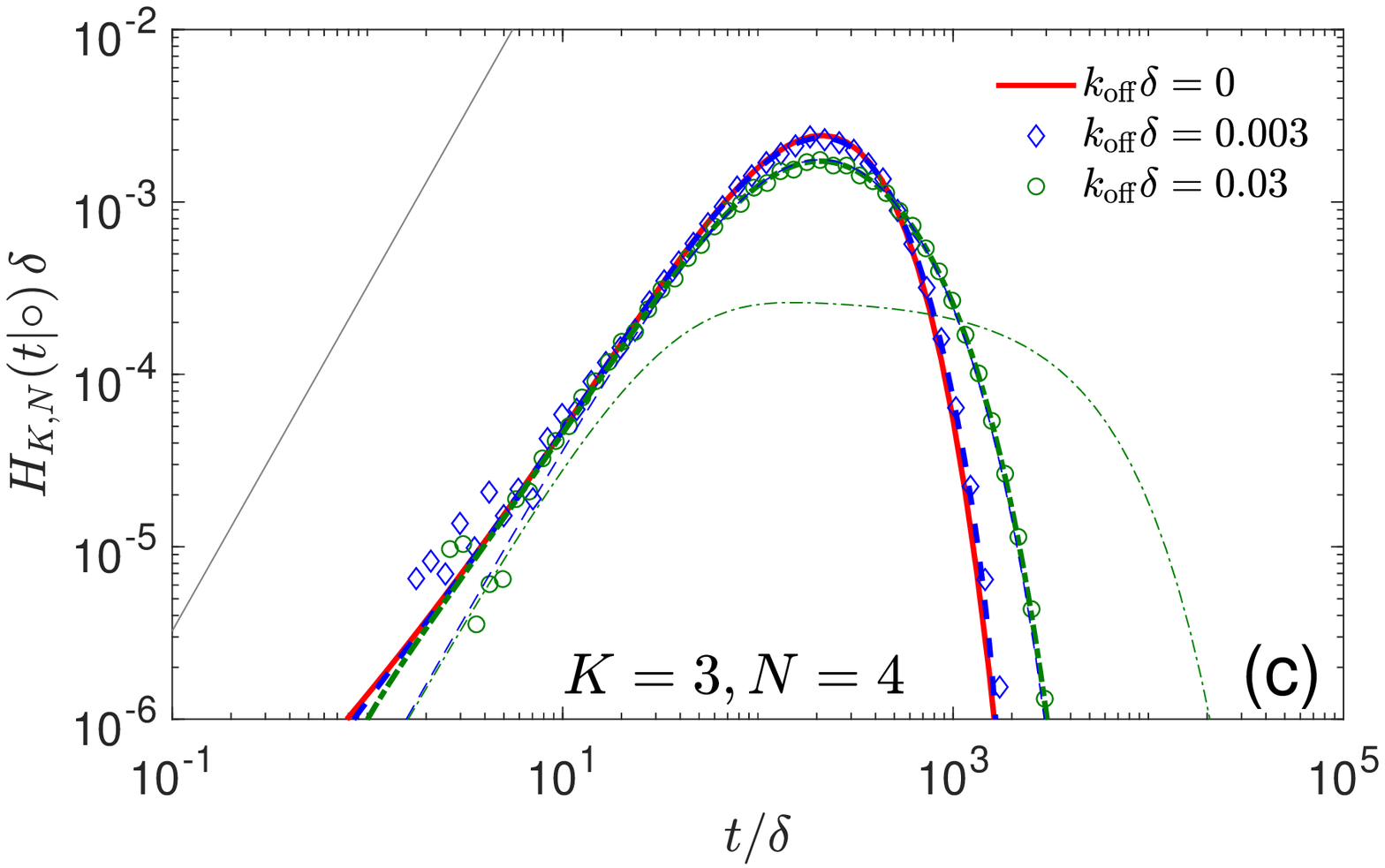} 
\includegraphics[width=77mm]{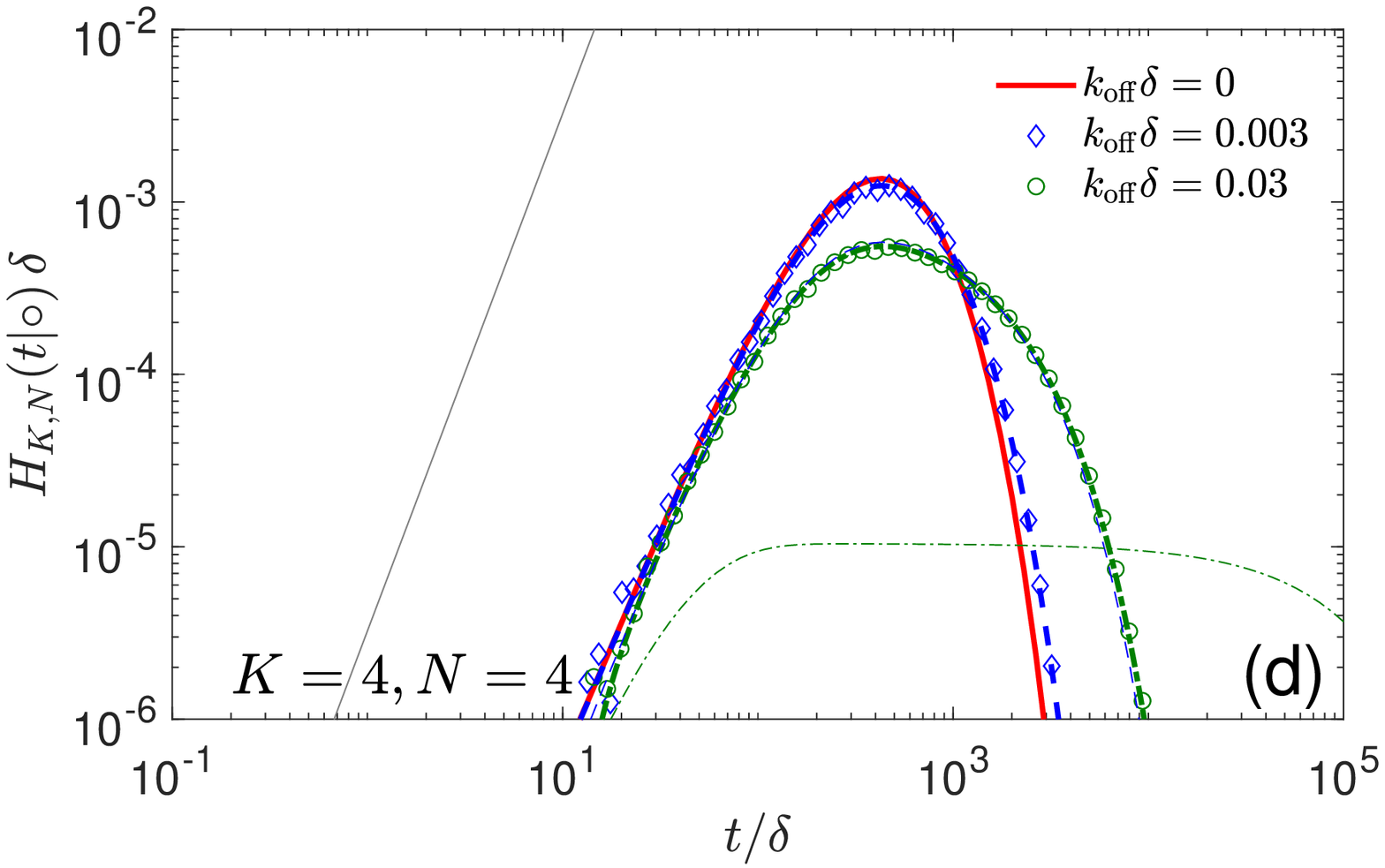} 
\end{center}
\caption{
Probability density of the reaction time $\T_{K,N}$ for restricted
diffusion between concentric spheres of radii $\rho$ and $R = 10
\rho$, with $N = 4$, $\kappa \rho/D = 10$, a timescale $\delta =
\rho^2/D$, three values of $\koff$ (see legend), and four values of
$K$: $K = 1$ {\bf (a)}, $K = 2$ {\bf (b)}, $K = 3$ {\bf (c)}, and $K =
4$ {\bf (d)}.  Symbols show empirical histograms from Monte Carlo
simulations with $10^6$ particles.  Thick lines indicate our
approximation (\ref{eq:HKN_exact}) evaluated numerically as described
in \ref{sec:computation}, whereas thin lines show the Lawley-Madrid
approximation (\ref{eq:HKN_LMA}), with $\nu$ given by
Eq. (\ref{eq:nu_LMA}).  Thin gray solid line presents the short-time
asymptotic behavior (\ref{eq:HKN_short}).  Minor deviations between
three thick curves on panel (a) at long times and on panels (c,d) at
short times can be related to insufficient discretization of
integrals, see \ref{sec:computation}. }
\label{fig:H4_kappa10}
\end{figure}

Figure \ref{fig:H4_kappa10} illustrates the behavior of the
probability density $H_{K,N}(t)$ for $N = 4$ and several values of $K$
when the target is highly reactive ($\kappa \rho/D = 10$).  One sees
that our approximation remains to be very accurate whereas the LMA
fails in this case. 

\begin{figure}
\begin{center}
\includegraphics[width=77mm]{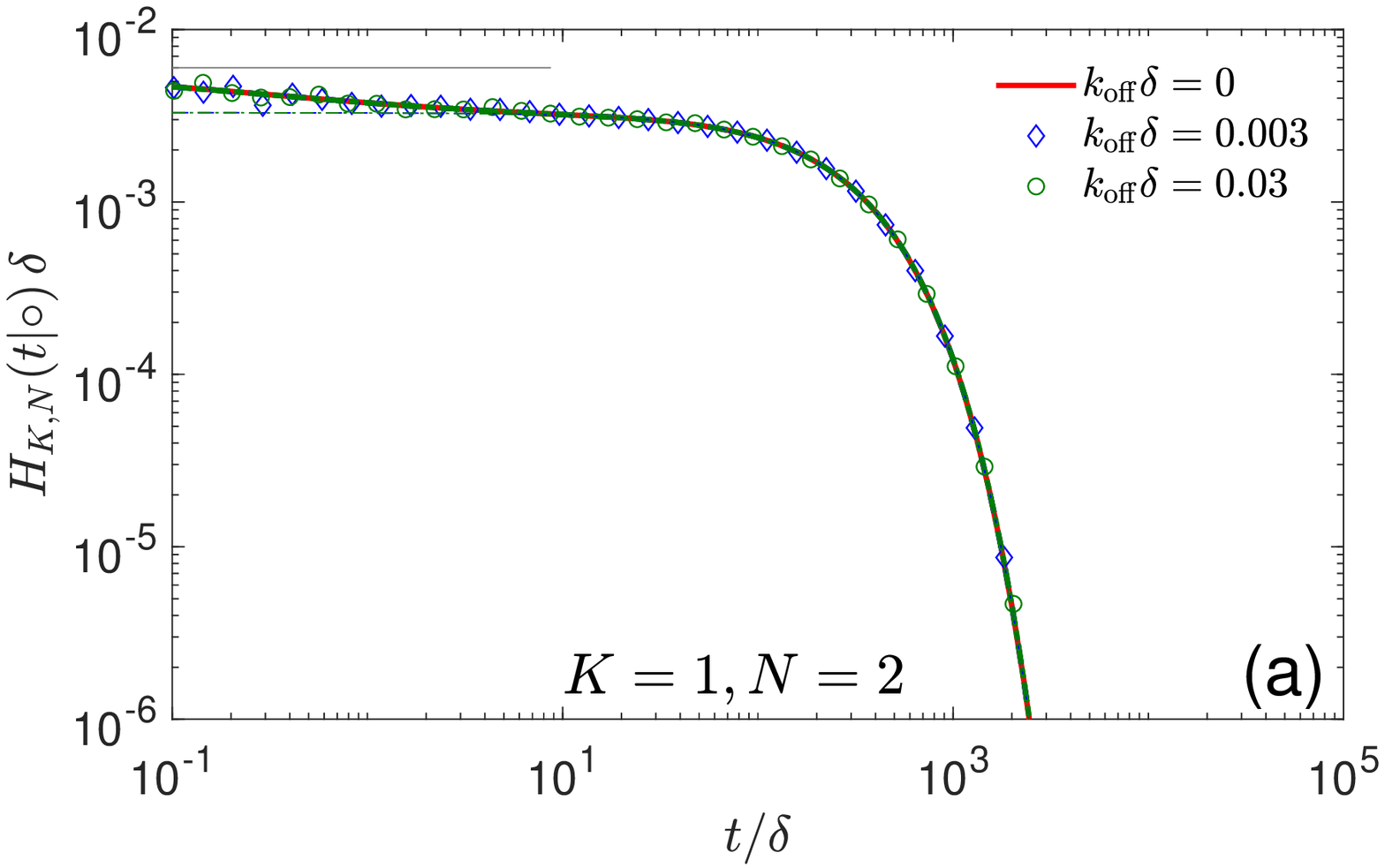} 
\includegraphics[width=77mm]{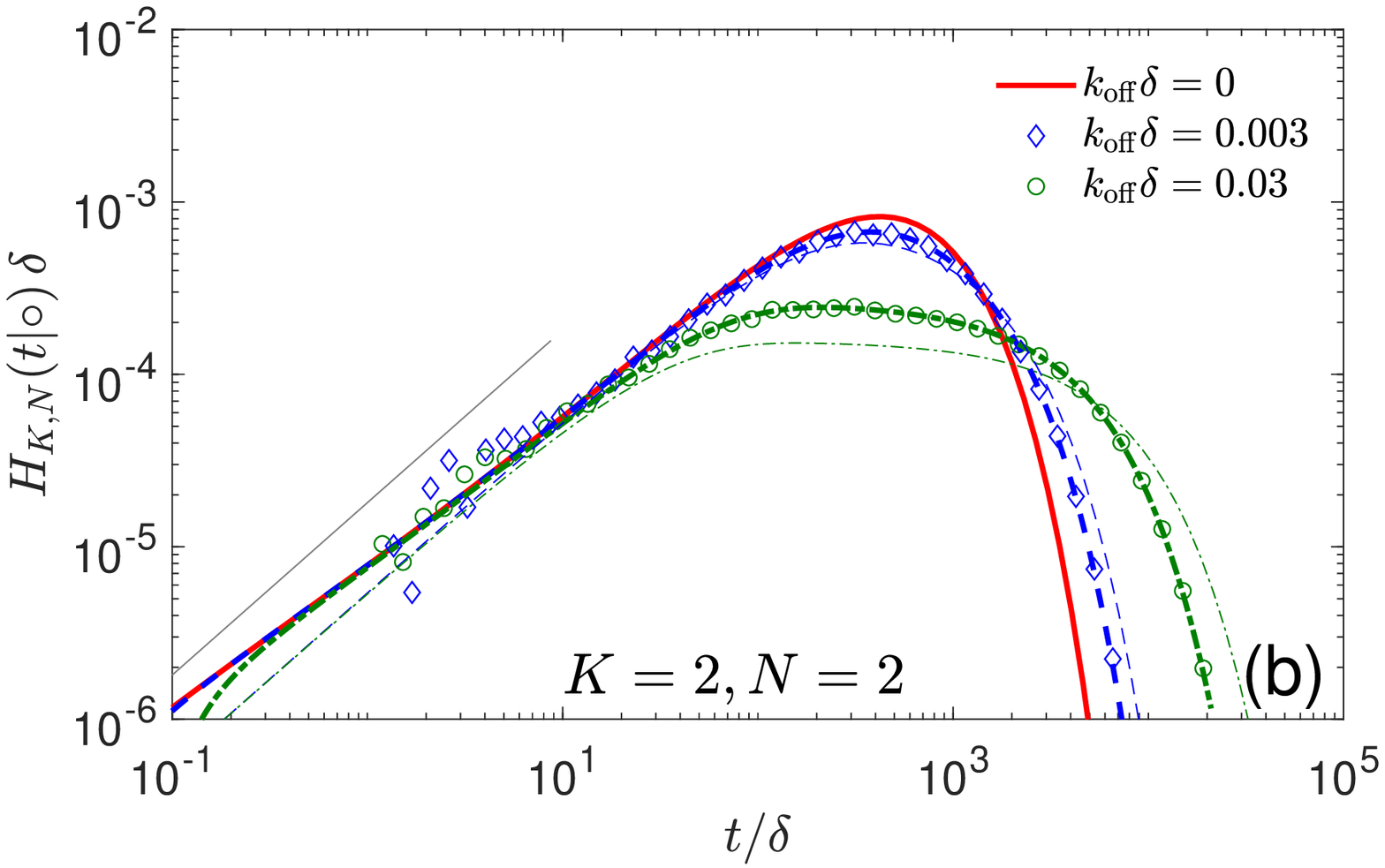} 
\end{center}
\caption{
Probability density of the reaction time $\T_{K,N}$ for restricted
diffusion between concentric spheres of radii $\rho$ and $R = 10
\rho$, with $N = 2$, $\kappa \rho/D = 1$, a timescale $\delta =
\rho^2/D$, three values of $\koff$ (see legend), and two values of
$K$: $K = 1$ {\bf (a)} and $K = 2$ {\bf (b)}.  Symbols show empirical
histograms from Monte Carlo simulations with $10^6$ particles.  Thick
lines indicate our approximation (\ref{eq:HKN_exact}) evaluated
numerically as described in \ref{sec:computation}, whereas thin lines
show the Lawley-Madrid approximation (\ref{eq:HKN_LMA}), with $\nu$
given by Eq. (\ref{eq:nu_LMA}).  Thin gray solid line presents the
short-time asymptotic behavior (\ref{eq:HKN_short}). }
\label{fig:H2}
\end{figure}

\begin{figure}
\begin{center}
\includegraphics[width=77mm]{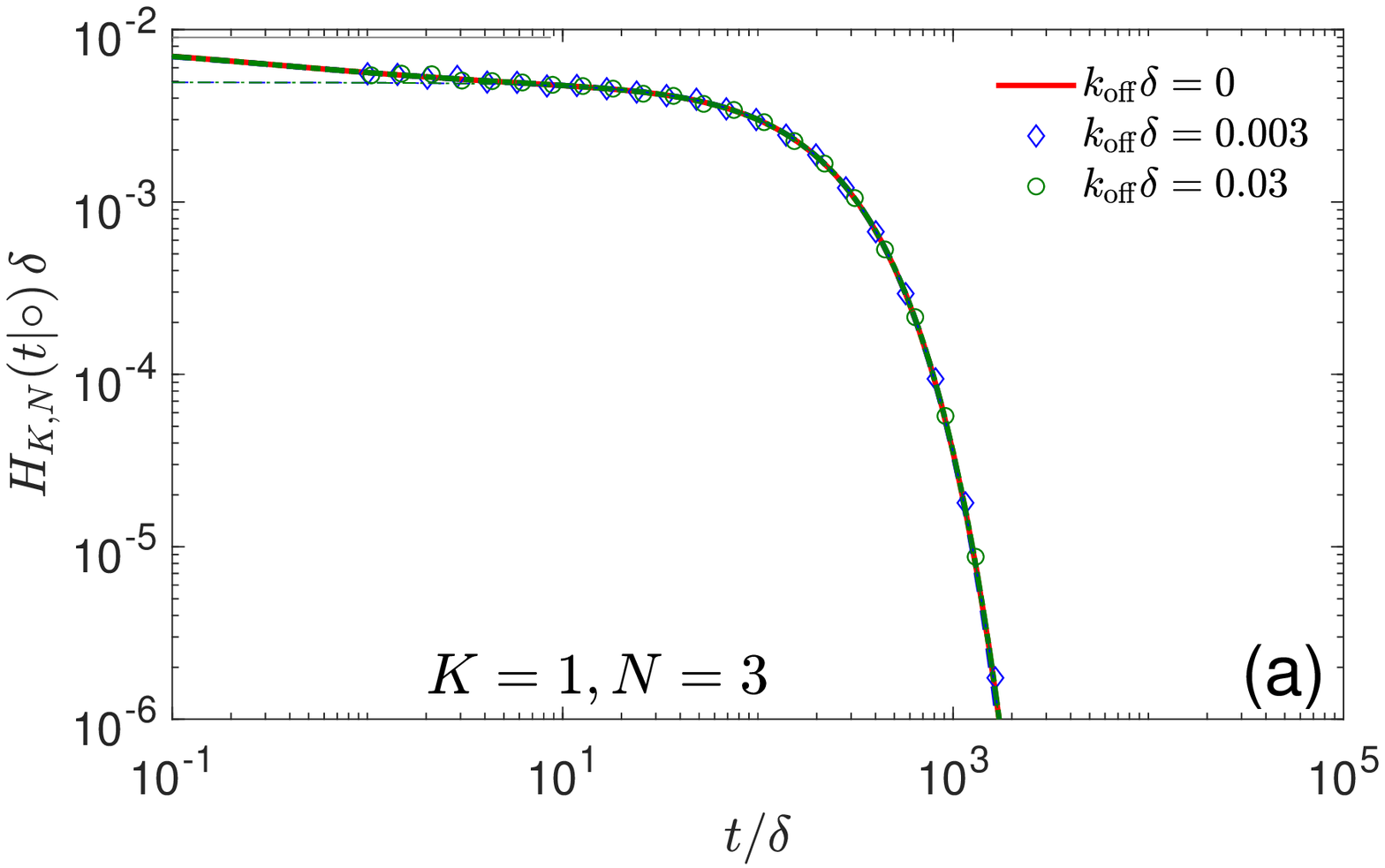} 
\includegraphics[width=77mm]{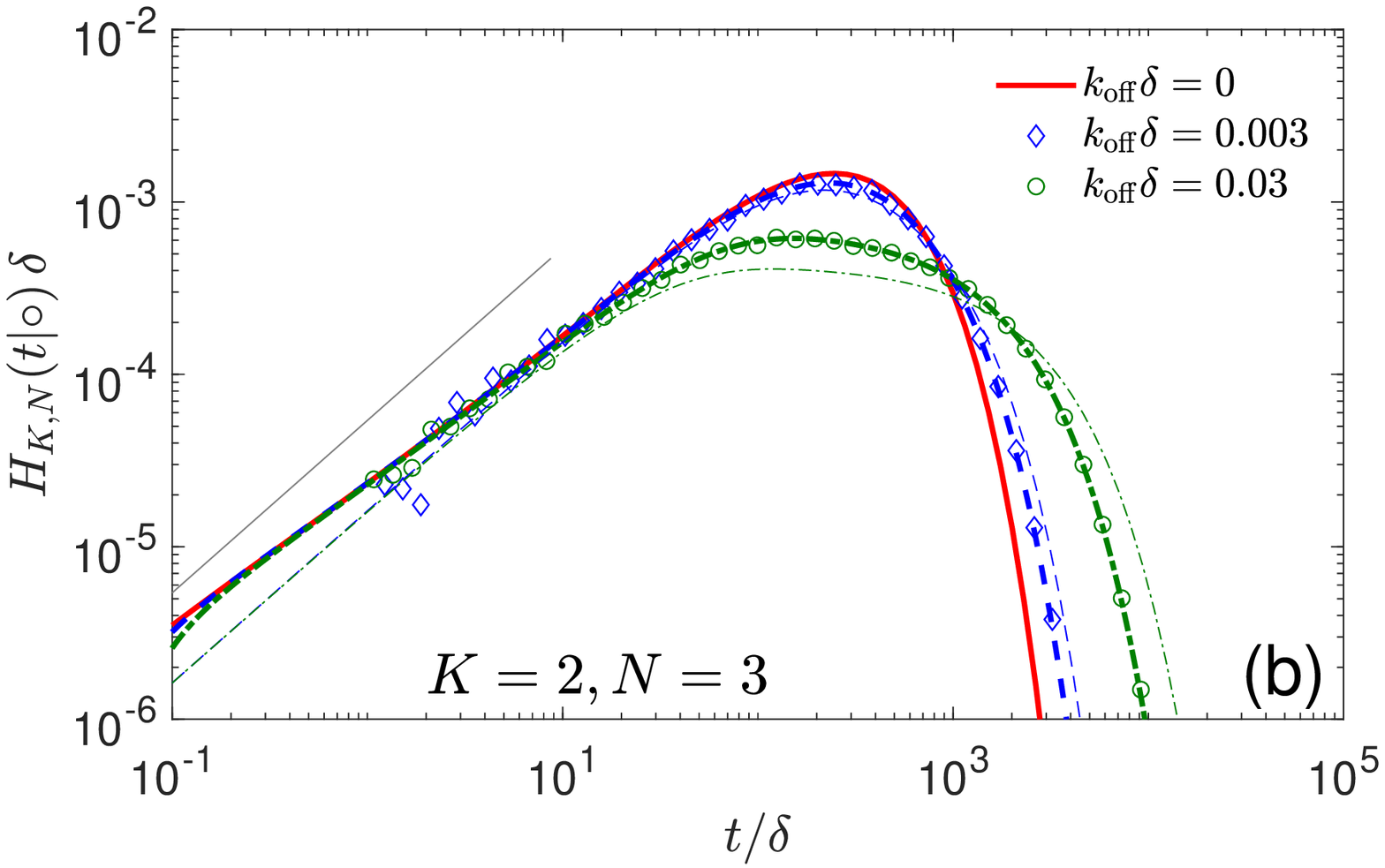} 
\includegraphics[width=77mm]{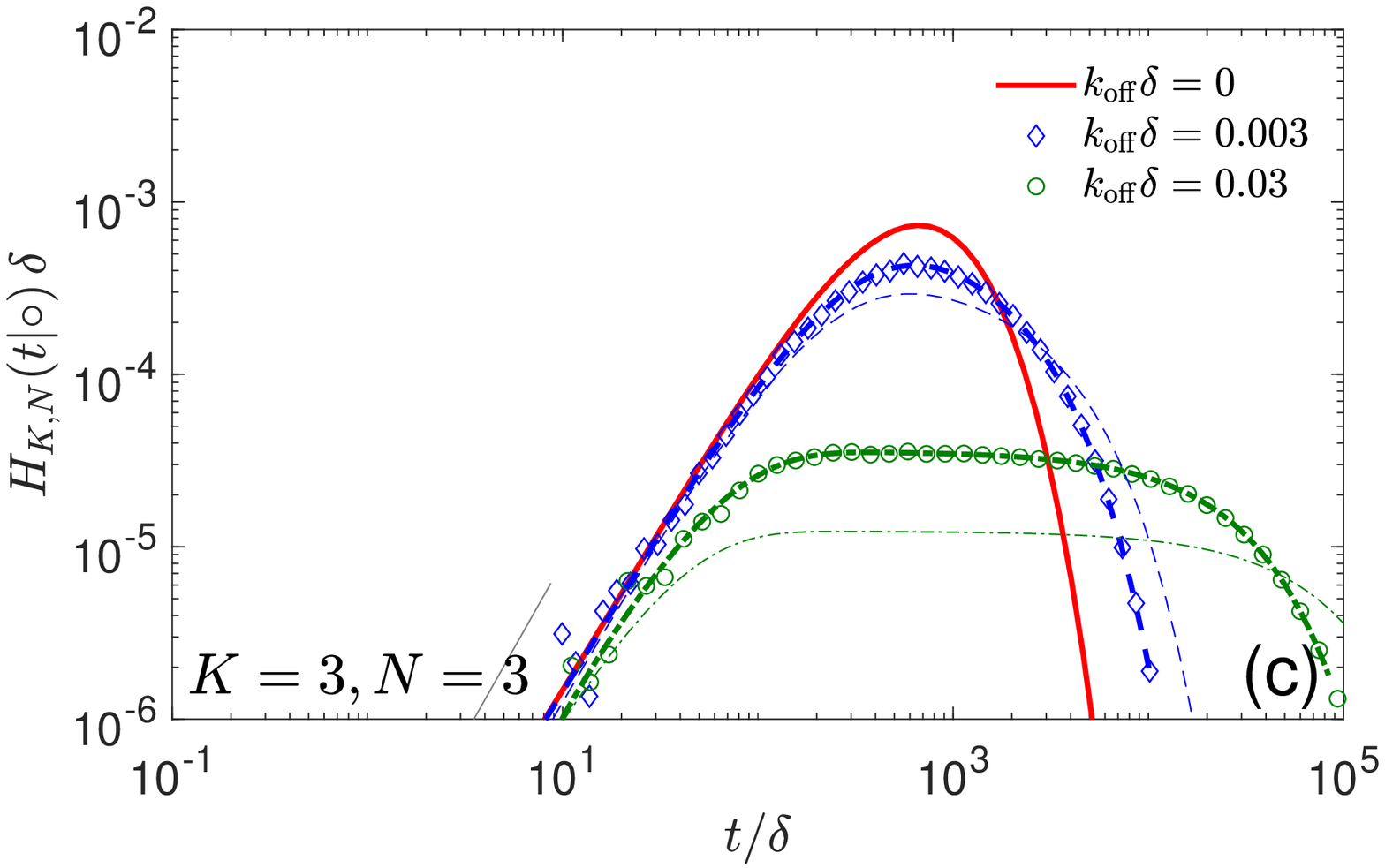} 
\end{center}
\caption{
Probability density of the reaction time $\T_{K,N}$ for restricted
diffusion between concentric spheres of radii $\rho$ and $R = 10\rho$,
with $N = 3$, $\kappa \rho/D = 1$, a timescale $\delta =
\rho^2/D$, three values of $\koff$ (see legend), and three values of
$K$: $K = 1$ {\bf (a)}, $K = 2$ {\bf (b)}, and $K = 3$ {\bf (c)}.
Symbols show empirical histograms from Monte Carlo simulations with
$10^6$ particles.  Thick lines indicate our approximation
(\ref{eq:HKN_exact}) evaluated numerically as described in
\ref{sec:computation}, whereas thin lines show the Lawley-Madrid
approximation (\ref{eq:HKN_LMA}), with $\nu$ given by
Eq. (\ref{eq:nu_LMA}).  Thin gray solid line presents the short-time
asymptotic behavior (\ref{eq:HKN_short}). }
\label{fig:H3}
\end{figure}

Figures \ref{fig:H2} and \ref{fig:H3} show the probability density
$H_{K,N}(t)$ for $N = 2$ and $N = 3$, respectively.  Its behavior is
similar to that discussed in the main text for Fig. \ref{fig:H4} with
$N = 4$.

\section{Further discussion on the validity of two approximations}
\label{sec:validity}

The Lawley-Madrid approximation relied on the assumption that both the
first-binding time and the rebinding time obey an exponential law with
some rate $\nu$, i.e., $S(t|\circ) \approx S(t) \approx e^{-\nu t}$.
In \ref{sec:mean}, we emphasized that the validity of our
approximation at small $\koff$ requires that $S(t|\circ) \approx
e^{-\nu t}$.  In this Appendix, we further discuss these points.

The spectral expansion (\ref{eq:So_spectral}) indicates that its
coefficients $a_n \geq 0$, defined by Eq. (\ref{eq:an}), can be
understood as the relative weights of different Laplacian eigenmodes,
given that $1 = S(0|\circ) = \sum\nolimits_{n=1}^\infty a_n$.  When
the target is small and/or weakly reactive, the ground eigenfunction
$u_1(\x)$ is almost constant so that $a_1 \approx 1$, whereas the
other eigenfunctions are orthogonal to it, implying $a_n \approx 0$
for $n > 1$ (see \cite{Lawley19,Grebenkov20}).  In other words, one
has $S(t|\circ) \approx e^{-\nu t}$, with $\nu = D\lambda_1$.  For
instance, when the target is a sphere of radius $\rho = 1$ surrounded
by a larger reflecting sphere of radius $R = 10$, we got numerically
$a_1 \approx 0.9989$ for $\kappa \rho/D = 1$ and $a_1 \approx 0.9946$
for $\kappa \rho /D = 100$, i.e., even for a highly reactive target,
the exponential law approximation is applicable for $S(t|\circ)$.
Even for a large highly reactive target with $\rho/R = 0.5$ and
$\kappa \rho/D = 100$, one has $a_1 \approx 0.92$, i.e., the ground
eigenmode still yields the dominant contribution.  This observation
justifies the high accuracy of our approximation even for highly
reactive targets.

It is also instructive to look at the parameter $\epsilon$ given
by Eq. (\ref{eq:LMA_cond}), whose smallness was required in
\cite{Lawley19} for the applicability of the Lawley-Madrid
approximation.  In our geometric setting, one gets $\epsilon =
\frac{\kappa \rho^2}{3DR(1+(\rho/R)^2)^2}$, so that $\epsilon \ll 1$
for $\rho/R = 0.1$ and $\kappa \rho/D = 1$, indicating the validity of
this approximation.  In contrast, $\epsilon$ is not small for other
examples given above thus violating the Lawley-Madrid approximation.

While the first-binding time can indeed be considered as exponentially
distributed, the situation is more subtle for the rebinding time
$\tau$ that is governed by the survival probability
\begin{equation}  \label{eq:St}
S(t) = \langle\tau\rangle H(t|\circ) = \sum\limits_{n=1}^\infty a_n D\lambda_n \langle\tau\rangle \, e^{-Dt\lambda_n} ,
\end{equation}
where we used Eq. (\ref{eq:Hcirc_S}).  The new coefficients $a'_n =
a_n D\lambda_n \langle\tau\rangle$ are as well the relative weights of
the eigenmodes.  Since the coefficient $a_1 \approx 1$ is multiplied
by a small eigenvalue $\lambda_1$, the resulting coefficient $a'_1$ is
not necessarily dominant.  For the above example with $\rho/R = 0.1$,
we get $a'_1 \approx 0.5474$ for a moderately reactive target ($\kappa
\rho/D = 1$), i.e., the contribution of the ground mode is still
dominant ($55\%$) but not exclusive.  In turn, for a highly reactive
target ($\kappa \rho/D = 100$), one has $a'_1 \approx 0.0119$, i.e.,
the contribution of the ground mode is only $1\%$.  In both cases, the
approximation of the rebinding time distribution by an exponential
distribution is not valid, and one needs much smaller or less reactive
targets to apply this approximation.  In summary, modeling the
rebinding time distribution by an exponential law imposes strong
restrictions onto the target size and reactivity.  As our
approximation employs the exact form of the probability density $H(t)$
of the rebinding time, it does not suffer from these limitations and
yields more accurate results than the Lawley-Madrid approximation.  In
turn, the latter has a great advantage of being much simpler and more
explicit.

The validity of the Lawley-Madrid approximation was discussed in
\ref{sec:LMA} and can be resumed by two inequalities (\ref{eq:LMA_cond2})
requiring that the target should be small and weakly reactive.  In
turn, quantitative conditions for the validity of our approximation
remain unknown.  In \ref{sec:mean}, we discussed a plausible condition
$\eta = \koff \langle\tau\rangle \gtrsim 1$, which can also be written
by using Eq. (\ref{eq:tau}) as
\begin{equation}  \label{eq:condition}
\kappa \lesssim \frac{\koff |\Omega|}{|\Gamma|} \,.
\end{equation}
For instance, for a small spherical target of radius $\rho$, it reads
\begin{equation}  \label{eq:kappa_upper}
\kappa \lesssim \frac{D}{\rho} (\koff T) ,
\end{equation}
where $T = |\Omega|/(4\pi D \rho)$ is the leading-order term of the
mean first-passage time to the perfect target from a starting point
uniformly distributed in $\Omega$ (alternatively, $1/(DT)$ is the
smallest eigenvalue of the governing Laplace operator, see
\cite{Mazya85,Ward93,Cheviakov11}).  This is a time scale of diffusive
search for a perfect target.  In turn, the second condition in
(\ref{eq:LMA_cond2}) for the applicability of the LMA imposes 
\begin{equation}  \label{eq:LMA_cond2a}
\kappa \ll D/\rho.  
\end{equation}
The comparison of these conditions illuminates the difference in the
validity ranges of two approximations.  In fact, when $\koff$ is not
too small (i.e., when $\koff T \gg 1$), the condition
(\ref{eq:kappa_upper}) is less restrictive than (\ref{eq:LMA_cond2a}),
and our approximation allows one to deal with highly reactive targets.
In contrast, it fails in the limit $\koff\to 0$, as illustrated in
\ref{sec:mean}, whereas the Lawley-Madrid approximation, whose
applicability is independent of $\koff$, can still be valid if
(\ref{eq:LMA_cond2a}) is satisfied.

We stress, however, that the conjectural condition $\eta \gtrsim 1$
and its equivalent forms (\ref{eq:condition}, \ref{eq:kappa_upper})
are not so restrictive in practice.  For instance,
Fig. \ref{fig:H4_comparison} shows a perfect agreement between our
approximation and Monte Carlo simulations in the case $\eta = 1$.  We
therefore expect that the range of applicability of our approximation
is much broader.  Its systematic study presents an important
perspective of this work.


\end{document}